 \documentclass[twocolumn,12pt]{revtex4}
\usepackage{mathtools}
\usepackage{amsmath,bbm}
\usepackage{amssymb,bm}
\usepackage{array}
\usepackage{graphicx}
\usepackage{enumerate}

\usepackage{slashed}
\usepackage[subcaption = false]{subfig}
\usepackage{caption}
\begin{document}
%\title{Charmonium suppression in QGP at LHC using temperature dependent recombination cross section}
%\title{Charmonium suppression in QGP at LHC using temperature dependent recombination cross section and formation time}
\title{pQCD approach to Charmonium regeneration in QGP at the LHC}
\author{S. Ganesh\footnote{Corresponding author:\\Email: gans.phy@gmail.com}}
\author{M. Mishra}
\affiliation{Department of Physics, Birla Institute of Technology and Science, Pilani - 333031, INDIA}
\begin{abstract}
  We analyze the applicability of perturbative QCD (pQCD) approach to the issue of $J/\psi$ recombination at the Large Hadron Collider (LHC), and calculate the recombination cross section for $c\bar{c}$ recombination to form $J/\psi$ as a function of temperature. The charmonium wavefunction is obtained by employing a temperature dependent phenomenological potential between the $c\bar{c}$ pair. The temperature dependent formation time of charmonium is also employed in the current work. 
A set of coupled rate equations is established which incorporates color screening, gluonic dissociation, collisional damping and recombination of uncorrelated $c\bar{c}$ pair in the quark-gluon plasma (QGP) medium. 
The final $J/\psi$ suppression, thus determined as a function of centrality is compared with the ALICE experimental data at both mid and forward rapidity and CMS experimental data at mid rapidity obtained from the Large Hadron Collider (LHC) at center of mass energy $\sqrt{s_{NN}} = 2.76$ TeV.  

\vskip 0.5cm

{\noindent \it Keywords} : Color screening, Recombination, Gluonic dissociation, Collisional damping, Survival probability, pQCD\\
{\noindent \it PACS numbers} : 12.38.Mh, 12.38.Gc, 25.75.Nq, 24.10.Pa 
\end{abstract}
\maketitle
\section{Introduction}
%gans rev1
    Charmonium and bottomonium (quarkonium) have been considered as important signatures to study the formation and properties of the quark gluon plasma (QGP). Various experimental~\cite{PKSMCA, PKSRAR, PKSAAD, ALICEfor,ALICEjpsi, PKSCMS1,PKSBAB} and theoretical work~\cite{PKSSPS1,PKSSPS2, PKSSPS3, PKSSPS4, PKSSPS5,PKSSPS6,PKSSPS7, yunpen,zhen,rishi} have been carried out to study the charmonium and bottomonium suppression. These signatures are similar in many respects, but differ mainly in the aspect of secondary charmonium/bottomonium formation i.e. recombination. Due to the heavy rest mass of the bottom quark, very little bottom quark and anti-quarks are formed even at the LHC energies. On the contrary, a relatively much larger number of charm quarks and anti-quarks are produced at the same energy, which can recombine to form charmonium even at the later stages of the QGP. There have been quite a few attempts for modeling the recombination phenomena to determine the effective quarkonium suppression.
%gans rev1
%Bottomonium suppression is considered to be a cleaner probe, since, even at the LHC energies, a sufficient number of secondary bottom quarks and anti-quarks could not be produced to reproduce $\Upsilon$ via recombination. As a consequence of this, one needs to only consider the suppression of the $\Upsilon$ formed due to the initial hard collisions. 

%But in the case of charmonium, due to lower rest mass of the charm quark, sufficient number of charm and anti-charm quarks are present in the later stages of QGP medium, which can recombine to form more number of charmonium states.   
%Therefore, a more precise estimation of $J/\psi$ suppression needs to incorporate the recombination mechanism with temperature dependent recombination cross-section which is the main aim of the current work. 
%Many authors incorporated the recombination of $c \bar c$ pair to quantify charmonium suppression at available collider energies by assuming a fixed value of recombination cross section at all the temperatures that ignores the Debye color screening effect during recombination and seems not to be very appropriate for explaining suppression data obtained from heavy-ion collision experiments. 
%The temperature dependent formation time~\cite{gans2} for charmonium states is another important ingredient of the present work.    

In Ref.~\cite{stathadref}, the authors have used statistical hadronization to capture the $J/\psi$ recombination, while in~\cite{twice1ref, trans2ref}, transport models have been utilized. 
In the approach adopted in Ref.~\cite{rituraj}, recombination cross section has been derived from the dissociation cross section using detailed balance approach. 
In this work, we look at the applicability of pQCD to calculate the recombination cross section. While recombination is a non-perturbative process, we provide arguments and simulation results that show that pQCD can be a fair approximation at the LHC energies. We also model the temperature dependence of the recombination cross section.

We model the recombination mechanism as a pair of uncorrelated charm quark and anti quark colliding to produce a gluon and a $J/\psi$. The Feynman diagrams corresponding to this mechanism are given in Appendix A. To validate the applicability of pQCD, we investigate the momentum transfer in the collision in the $c\bar{c}$ center of mass frame of reference.
In the diagrams $1$ to $5$, it can be seen that the uncorrelated $c$ and $\bar{c}$ pair annihilate to form a virtual gluon, which then creates a correlated $c\bar{c}$ pair and eventually forms a $J/\psi$. The minimum energy of the incoming $c\bar{c}$ pair has to be equal to the $J/\psi$ rest mass $=3.1$ GeV. In the center of mass frame, the net input momentum is $0$, which implies that the virtuality of the gluon propagator needs to be $\ge$ $J/\psi$ rest mass ($3.1$ GeV). This is within the realm of pQCD.
In the diagrams $6$ to $10$, there is no annihilation of incoming $c\bar{c}$ pair, but there is a momentum transfer via the virtual gluon. The momentum of the bound quark and anti-quark in $J/\psi$ and the momentum of the outgoing gluon is transferred via the gluon propagator. 
%If the different between the $J/\psi$ rest mass and $c\bar{c}$ rest mass is attributed to the average momentum of the charm quark and anti-quark, a momentum transfer of about 0.88GeV happens via the virtual gluon, leading to a virtuality of 0.88GeV. This borders on $\Lambda_{QCD}\,\approx\,1GeV$. 
From the solution of the Schr\"{o}dinger equation, it can be seen that the standard deviation of the momentum spread of the bound charm quark and anti-quark is about $0.5$ GeV and this helps in reducing the error due to pQCD even if the emitted gluon is not very hard. The distribution of the momentum of the outgoing gluon plays a significant role in determining the error due to pQCD.
	
Given the large $p_T$ range of many (though not all) of the outgoing $J/\psi$ at the LHC, many of the gluons are likely be hard gluons. The accuracy of pQCD calculations would depend upon the percentage of hard gluons emitted.
 In the initial hard production, the correlated $c$ and $\bar{c}$ are generated from the colliding nucleons, with the collision axis being along the beam axis. 
However, in the case of the formation of secondary $J/\psi$ produced from recombination of the uncorrelated charm quark, and anti quark, the collision axis need not be directed along beam axis. Consequently, the $p_T$ of the emerging $J/\psi$ need not necessarily determine the hardness of the process. 
For instance, a pair of particles moving almost in parallel may undergo a soft collision, and could be Lorentz boosted due to their perpendicular (to the collision axis) momentum, resulting in a large $p_T$. 
But, since the collision axis and the $J/\psi$ emission can be directed in any random direction, the effect of Lorentz boost can be in either direction i.e., to either increase or decrease $p_T$. 
Hence, though the $p_T$ of the outgoing $J/\psi$ (gluon) in the lab. frame is not an exact measure of hardness, it can act as a probabilistic estimate for the momentum range of $J/\psi$ (gluon) in center of mass frame and thus can be an approximate measure of the hardness of the gluon emitted. In other words, a large momentum range in the lab. frame is likely to result in a large momentum in the center of mass frame.
For Pb$-$Pb collision, at the ALICE mid rapidity, the $p_T$ range is $0<p_T<8$ GeV, while at the ALICE forward rapidity, the $p_T$ range is $p_T > 0$. At the CMS mid rapidity, $p_T$ range is $6.5 < p_T<30$ GeV. 
At forward rapidity, the longitudinal component of the momentum  is significant, and can significantly increase the total magnitude of $J/\psi$ momentum compared to ALICE mid rapidity. We describe this more quantitatively in the latter part of the  introduction.
In the light of the foregoing discussion, the applicability of pQCD may increase in the order: 
\\
ALICE(mid-rapidity) $<$ ALICE(forward rapidity) $<$ CMS(mid-rapidity,).\\ 
In fact, from our simulation results, shown in section IV, it can be seen that the experimental $J/\psi$ suppression data is reproduced to a much better extent for ALICE forward rapidity and CMS mid rapidity than at ALICE mid rapidity. 

  The last vertex that needs to be explored is the outgoing gluon. The outgoing gluon can be a soft gluon, but as argued above, if the number of hard gluons are large, the discrepancy due to application of pQCD can be small. 
Secondly, in Appendix B, we show that the amplitude due to the soft gluon can partially cancel between the various pairs of diagrams (Eg., between diagram $6$ and $7$ or $2$ and $3$ etc.). If the soft gluon is perpendicular or parallel to the collision axis, then it exactly cancels, and for other inclinations, it cancels partially to varying degrees. 
%It is to be noted that the perpendicular gluon has the maximum phase space (spans the largest circle around the collision axis) and thus provides the maximum contribution to the cross section. 
   This, we expect, would soften the divergence of the Fermion propagator, and reduce the error due to divergence of the Fermion propagator in the pQCD calculations.
   To further validate all the above arguments for the approximate validity of pQCD, we did numerical simulation in which the  amplitudes with gluon virtuality $<~1$ GeV or outgoing gluon momentum $<~1$ GeV were zeroed out. The resulting recombination rate is indicated by the dashed line in Fig. 2. Comparing the difference between the solid and the corresponding dashed curves, the recombination decreases by a maximum of about 9\% at ALICE mid rapidity and by a much smaller extent at ALICE forward rapidity. 
The difference is indicative of the number of soft gluons present. 
This corroborates with the above arguments on the gluons being mostly hard. 
To summarize, while an exact treatment of $J/\psi$ recombination would require a non-perturbative analysis, the arguments and simulation results indicate sufficient grounds to believe that pQCD provides a fair approximation of the recombination mechanism at LHC energy. 

At ALICE forward rapidity, in order to estimate the longitudinal component, we analyze the total momentum for $p_T \ge 0.35\,$ GeV. At $p_T = 0.35\,$ GeV, the longitudinal momentum $p_l \ge 2.15$ GeV for the rapidity range $2.5 < y < 4.0$. 
The  total $J/\psi$ momentum becomes around $2.2$ GeV, corresponding to energy $= 3.8$ GeV. 
The number of $J/\psi$ particles less than $p_T <  0.35$ GeV may be small. As a supporting evidence, from Ref.~\cite{pp7TeV}, the $J/\psi$ production peaks at around $p_T = 1.8$ GeV at forward rapidity $\sqrt{S_{NN}} = 7$ TeV. Assuming the distribution to be similar for $Pb-Pb$ collision at $\sqrt{s_{NN}} = 2.76$ TeV, only a small number of $J/\psi$ particles will lie in the range $p_T < 0.35\,$ GeV. Additionally, there would be $J/\psi$ particles with $p_T < 0.35$ GeV, but still with total momentum greater than $2.2$ GeV. 
With the charm quark distribution (solid line) in Fig. 1, an estimate of the number of $J/\psi$ particles with total momentum $\ge$ $2.2$ GeV is $99$\%. 
%With the modified Fermi Dirac distribution ($\frac{N}{exp(-\lambda E/kT) + 1)}$),which we shall describe in section IV, it is 93\%,
Hence, for ALICE forward rapidity, we take $2.2$ GeV as the lower momentum cut off.

%At this point, we would also like to  mention that if a final outgoing gluon is also present (which is what we consider in this work), the outgoing gluon contributes an additional 2.2GeV in the $c\bar{c}$ center of mass frame of reference.  This lends support to the applicability of pQCD arguments we had made earlier. 

%Thus, in the process of limiting $p_T$ to $0.35\,GeV$, we are ignoring the suppression of only a small number of $J/\psi$.  
%For the measurements done at forward rapidity at ALICE with $p_T > 3\,GeV$\cite{ALICEfor}, recombination would be expected to be negligible.

%It is to be noted that the quantity which enters the rate equation is $\Gamma_{recomb}$ and not $\sigma_{recomb}$.
%  $\Gamma_{recomb} = \sigma_{recomb} \times v_{rel} \times fp(c)$, where $fp(c)$ is the charm quark distribution. 
%Furthermore, from Fig. 1, the charm quark population reduces drastically below energy = 2.8GeV. It falls 5 times from 2.8GeV to 1.5GeV. This implies that below charm quark anti quark pair energy of 5.6GeV, the number of charm quark pairs decreases drastically.
%Recombination depends quadratically on the number of charm quarks, and hence the number of recombination taking place below 5.6GeV should be very small.  
%gans rev1 rituraj recomb
In Ref.~\cite{rituraj}, the recombination cross section has been derived from the dissociation cross section~\cite{Wolschin} using detailed balance approach. 
The dissociation cross section developed in~\cite{Wolschin} is an extension over~\cite{peskin} by including confining string contribution using NRQCD. However, this formulation has been developed for soft gluons. One would thus expect it to be accurate for gluon momentum up to $\approx$ $1.0$ GeV or $1.5$ GeV.
%The dissociation cross section is based on the formulation given in ~\cite{Wolschin}, which has been developed for soft gluon emission ($\le 1.5$ GeV approx.).
As a result, the resulting recombination cross section would only be a rough approximation when applied to hard processes, especially at the ALICE forward rapidity ($p \ge 2.2$ GeV) and CMS mid rapidity ($p_T \ge 6.5$ GeV). The current pQCD formulation may be a better approximation at both ALICE forward rapidity and CMS mid rapidity. Even at ALICE mid-rapidity sufficient number of gluons would be expected to have $p_T \ge 1.5$ GeV. A comparison is drawn with the results of~\cite{rituraj} at the CMS mid rapidity in section~\ref{sec:results}. 
The second main difference from the model presented in ~\cite{rituraj} is the application of temperature dependent formation time which is described later in section~\ref{sec:rev1:color}.
%gans rev1 rituraj recomb

We use the pQCD calculations to determine the temperature dependent recombination time constant, $\Gamma_{recomb}(T)$ which is based on the temperature dependent cross section $\sigma^T_{recomb}$ for $c\bar{c}~\rightarrow~J/\psi$ process at various energies of the $c\bar{c}$ pair. The charmonium wavefunction used in the  calculation of $\sigma^T_{recomb}$ is determined by solving Schr\"{o}dinger wave equation with a temperature dependent Debye color screened phenomenological potential. 
As mentioned earlier, we mainly consider the process of a quark and anti-quark annihilation to form $J/\psi$ and a gluon in the final state, which is an $\alpha^3$ process. In the $\alpha^2$ process, where there is no final state gluon, the $c\bar{c} \rightarrow J/\psi$ reaction would involve only a very limited phase space domain, in which the sum of energy of incoming $c\bar{c}$ pair needs to be exactly equal to the $J/\psi$ rest mass in the $c\bar{c}$ center of mass frame of reference. Due to this extremely limited phase space domain, it plays an insignificant role in the recombination process. The other $\alpha^3$ processes involving gluon, quark and anti-quark in the initial state have similar limited phase space restrictions. Moreover, they involve three body collision and these processes are ignored. 
   The $J/\psi$ dissociating into a $c\bar{c}$ pair, and $c\bar{c}$ recombining to produce $J/\psi$ form a system of coupled rate equations, with each feeding into the other. In this work, we model this phenomenon by a system of coupled rate equations. 

   Furthermore, $J/\psi$ and $c\bar{c}$ would be formed from the interaction of light quarks and gluons in the medium. The pQCD framework developed here has used a relativistic modeling of the initial $c\bar{c}$ pair, and hence can be utilized to calculate the temperature dependent cross section for the lighter quarks to form $J/\psi$. In section II (D), we analyze the impact of the light quarks. Additionally, some of the $c\bar{c}$ would also decay back to the lighter quarks and gluons. We ignore this decay process in the current work. It is to be noted that the process involving the lighter quarks, would not be part of the feedback mechanism happening between $J/\psi$ and $c\bar{c}$, and hence, they would have a much lesser impact to the overall suppression or enhancement (via recombination) as compared to the $c\bar{c} \leftrightarrow J/\psi$ process. 
We also ignore the running of the coupling constant, and instead use a fixed value of the coupling constant.

   It is found that the charmonium suppression in the QGP is not the consequence of a single mechanism, but is a complex interplay of multiple mechanisms like color screening, gluonic dissociation, collisional damping and recombination. We develop a framework of rate equations, which incorporates all the above mechanisms. Color screening is based on quasi-particle model (QPM) equation of state (EOS) of the QGP described in~\cite{Madhu2}. The QGP is expanding under Bjorken's scaling law, applicable mainly at mid rapidity. Due to the modification of the formation time by temperature~\cite{gans2}, we show that the color screening becomes very small. Hence, we ignore color screening to compute suppression at forward rapidity. This could lead to a little under suppression in peripheral collisions. The process of gluonic dissociation and collisional damping is based on the formulation developed by Wolschin~\cite{Wolschin}. At the LHC energies, absorption is expected to be negligible, since the $c\bar{c}$ pair would behave almost as a color singlet while traversing through the nucleus and hence have negligible interaction with the nucleus. The $p_t$ broadening due to Cronin effect is inconsequential while explaining the $p_t$ integrated suppression at the mid rapidity for various centrality bins. 
That is why, we incorporate only shadowing based on the work done by Vogt~\cite{vogt} as a CNM effect in our present calculation. In fact, in Ref.~\cite{vogt}, the shadowing for Pb$-$Pb collisions at LHC energy $\sqrt{s_{NN}}=5.5$ TeV was determined. We use the same framework to calculate shadowing at $\sqrt{s_{NN}}=2.76$ TeV energy. 
%We do not however include the initial state collisional and radiative energy loss in CNM, which can affect nuclear modification factor, $R_{AA}$~\cite{cnmel0,cnmel1,cnmel2}. 
Our results for $J/\psi$ suppression versus $N_{part}$ are compared with the experimental data in the forward and mid rapidity region obtained from the ALICE~\cite{ALICEfor,ALICEjpsi} and mid rapidity data from CMS~\cite{PKSCMS1} experiments at LHC energy $\sqrt{s_{NN}}=2.76$ TeV. We find that our predictions show a reasonably good agreement with the suppression data, particularly at ALICE forward rapidity and CMS mid rapidity. 
%We also compare the results with the suppression due to fixed recombination cross section without color screening effect for the  momentum range $> 2.2\,GeV$ with  $p_T > 0.35  GeV$. 
%From now onwards, we refer to the momentum range $ \ge 2.2 GeV $,  as the ALICE mid momentum range, $0$ GeV to $8$ GeV as the ALICE low momentum range and $6.5$ GeV to $30$ GeV as the CMS momentum range. 
 
   The organization of the rest of the paper is as follows. Section II introduces the system of coupled rate equations, and describes the pQCD calculation of the temperature dependent recombination cross section. Section III briefly describes the mechanism of color screening, gluonic dissociation, collisional damping and CNM effects. The last part of section III, describes the inclusion of all these mechanisms in the rate equations. The section IV gives the results and discussions and finally conclusion is given in section V.  

\section{Rate Equations and the Recombination Cross Section}
\subsection{The coupled rate equations}
Once the charmonium bound state is formed after the formation time, we assume that there are two reversible processes which are in play. The first one is the dissociation of charmonium into its constituent charm and anti-charm quark. The second is the recombination of the charm and anti-charm quark to again form charmonium.
For a QGP with instantaneous volume $V(t)$, we model these two processes using the following set of rate equations:
\begin{eqnarray} 
\frac{dN_{J/\psi}}{dt} = -\Gamma_{diss}(T) N_{J/\psi}(t) + \Gamma_{recomb}(T) \frac{N_c(t) N_{\bar{c}}(t)}{V(t)} \\ \nonumber
\frac{dN_c}{dt} = \Gamma_{diss}(T) N_{J/\psi}(t) - \Gamma_{recomb}(T) \frac{N_c(t) N_{\bar{c}}(t)}{V(t)},
\end{eqnarray} 
where $\Gamma_{diss}$ and $\Gamma_{recomb}$ are $J/\psi$ dissociation and $c\bar c$ recombination rate, respectively. 
The above equations are solved numerically. 
The initial conditions are given by $N_{J/\psi}(0)$, $N_c(0)$ and $N_{\bar{c}}(0)$ after including shadowing. $N_c(0)$ and $N_{\bar{c}}(0)$ are determined from~\cite{thews} and taken as $75$ based on the MRST HO structure function of nucleon for the most central collision. The ratio of $N^{pp}_{J/\psi}$ to $N^{pp}_c$ is determined from the experimental data~\cite{PKSCMS1,ccbar}. 
The value of $N^{pp}_{c}$ and $N^{pp}_{\bar{c}}$ is obtained from the open charm $D$ meson data from LHC~\cite{ccbar} and extrapolated till $p_t = 30$ GeV using the techniques mentioned in~\cite{interpolate, comov}. 
The open charm $D$ meson distribution is then scaled along the momentum axis according to the momentum fraction of the charm quark present in the $D$ meson to obtain the charm quark momentum distribution. The resultant normalized charm distribution is shown in Fig. 1 (solid line).
 The extrapolated value of $N_c$, after adjusting for the difference in luminosity for $N^{pp}_{J/\psi}$ and $N^{pp}_{c}$ data, is then used to determine the ratio $\frac{N^{pp}_{J/\psi}}{N^{pp}_c}$. This ratio is then used to obtain $N_{J/\psi}(0)$ from $N_{c}(0)$. The obtained values of $N_c(0)$ and $N_{J/\psi}(0)$ are then scaled for various centrality bins according to $T_{AA}$ and shadowing to arrive at the initial conditions for the rate equations. Modeling of the shadowing is described in the section on CNM effects.

%gans rev1 rate equation
In view of the fact that the initial number of $J/\psi$ is negligible, the coupled rate equations can be simplified leading to an analytical solution as outlined in~\cite{rituraj}. The difference between the numerical solution of the coupled rate equation and the analytical solution of the simplified equations is small. 
In this work, we have used the coupled rate equations for its more generic and allows the analysis of other effects, for example, the effect of light quarks in section~\ref{sec:light}. It also helps in more accurately analyzing the discrepancy in $J/\psi$ yield due to uncertainty in the initial conditions as discussed in Section~\ref{sec:results}.
\begin{figure}[h!]
%\vspace{2in}
\includegraphics[width = 80mm,height = 80mm]{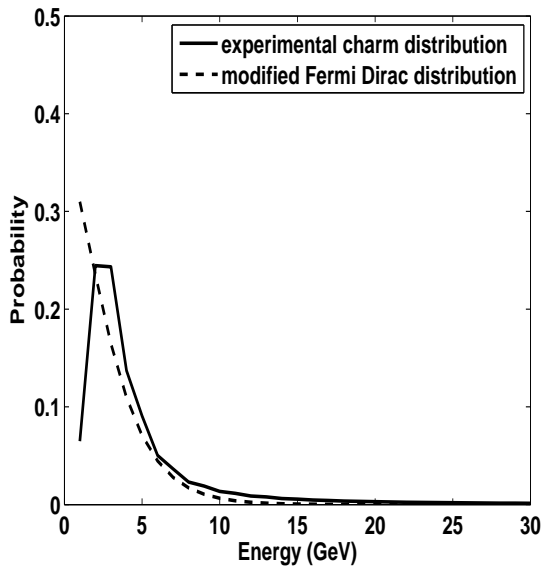}
\captionsetup{justification=raggedright, singlelinecheck=false}
\caption{Normalized Distribution of charm quark}
\label{fig:charm_distribution}
\end{figure}

The value of recombination time constant (recombination reactivity) is given by
$\Gamma_{recomb}(T) = \int dp_1 dp_2\,f_c(p_1)\,f_{\bar{c}}(p_2)\,\sigma_{recomb}(p_1,p_2)\,v_{rel}$, where $v_{rel}$ is the relative velocity between $c$ and $\bar{c}$ in the laboratory frame. 
$f_c(p_1)$ and $f_{\bar{c}}(p_2)$ are distribution function of $c$ and $\bar{c}$ with four momenta $p_1$ and $p_2$, respectively.
$\sigma^T_{recomb}$ is computed using pQCD calculations.
The suppression or enhancement modeled by the rate equations is defined by
%\begin{equation}
%	S_{dr} = \frac{N_{J/\psi}(\infty)}{N_{J/\psi}(0)}
%\end{equation}
%In practice, we define $S_{dr}$ as 
\begin{equation}
	S_{dr} = \frac{N_{J/\psi}(t_{QGP})}{N_{J/\psi}(t_0)},
\end{equation}
where $t_{QGP}$ is the lifetime of QGP ($=5$ fm) at the LHC energy, and $t_0$ is some initial time, which is taken to be the thermalization time $= 0.5$ fm.
\subsection{pQCD calculation of $\sigma_{recomb}$}

We use pQCD to compute the cross section of charm quark and anti-charm quark to form a final state gluon and $J/\psi$. 
%gans rev1 Feynman
In Ref.~\cite{QQanni}, Feynman diagrams for $Q\bar{Q}$ scattering to order $\alpha^2$, and for $Q\bar{Q}$ annihilation (production) to (from) light quarks and gluons are given. Since recombination involves production of $J/\psi$ from uncorrelated $c\bar{c}$ pair (instead of light quarks and gluons), the Feynman diagrams in~\cite{QQanni} need to be modified.
%gans rev1 Feynman
The Feynman diagrams corresponding to the recombination process are given in the Appendix A.  
We treat the $J/\psi$ bound state as a linear superposition of definite momentum eigenstates of a free particle. The contribution of each momentum eigenstate is given by the wavefunction $\psi_T(k,k^{\prime})$.
The charmonium bound state is modeled non-relativistically (a solution of a Schr\"{o}dinger equation)~\cite{rituraj, gans}, with the charm quark and anti-charm quark assumed to have very small momentum spread. From our solution to the Schr\"{o}dinger equation at the minimum QGP temperature of $170$ MeV, i.e. before hadronization, the $J/\psi$ wavefunction is seen to have a one standard deviation spread of $0.5$ GeV (approx.). At higher temperatures, the wavefunction expands spatially leading to a decrease in the momentum spread.

The four momentum conserving delta function $\delta^4(p_1 + p_2 - k_g - k - k^{\prime})$ would then imply that the momentum spread of bound state charm quarks and anti-charm quarks needs to be compensated by the momentum spread of input quarks and outgoing gluon. The incoming quarks and anti-quarks may naturally acquire a momenta spread due to confinement in QGP medium. 
The remaining needs to be compensated by the gluon, which may make the gluon go off shell by a small amount ($< 0.5$ GeV). 
%At higher non zero temperatures of the QGP, where the momentum spread of the bound $c\bar{c}$ pair is lesser, the gluon needs to go off shell by a much smaller amount than $0.35$ GeV. 
%For the more peripheral bins, the momentum spread of the bound quarks is lesser, and hence the gluon needs to go off shell by a much lesser amount.
%In the current modeling, $0.5$ GeV is substantially small compared to an average energy of $10.3$ GeV and the $p_t$ range of $6.5$ GeV to $30$ GeV.
%Comparison with the lower limit of $3.8$ (=$\sqrt(3.1^2 + 2.2^2)$) GeV at ALICE forward rapidity conditions, $0.5$ GeV is still small. 
As an approximation, we ignore the effect of the momentum spread of the incoming charm quarks, anti-charm quarks and the outgoing gluon and place the gluon fully on shell.
This could however be a source of error for soft gluons emitted particularly at ALICE mid rapidity. 

With the above approximation of the input quark being in a definite momentum eigenstate and the gluon being on shell, the probability amplitude for the various Feynman diagrams (given in Appendix A) is given by

\begin{enumerate}
\item 
\begin{math} 
%This mu and nu is as in my code
%\nonumber M_1 = \bar{v}(p_2)ig\gamma_{\mu}t^a u(p_1)(\frac{-i}{k_1^2}.gf^{abc})\\ \nonumber \left [g^{\mu\nu}(k_1 - k_g)^{\sigma} + g^{\nu\sigma}(k_g - k_2)^{\mu} + g^{\sigma\mu}(k_2 - k_1)^{\nu})\right ] \\ \frac{-i}{k_2^2}\epsilon_{\nu}^b(k_g)\bar{u}(k)(ig\gamma_{\sigma}t^cv(k'))
%interchanging nu and mu to be consistent with other equations.
%1
\nonumber M_1 = \Big ( \bar{v}(p_2)ig\gamma_{\nu}t^a u(p_1)\Big ) (\frac{-i}{k_1^2}.gf^{abc})\\ \nonumber \Big [ g^{\nu\mu}(k_1 + k_g)^{\sigma} + g^{\mu\sigma}(-k_g - k_2)^{\nu} + g^{\sigma\nu}(k_2 - k_1)^{\mu})\Big ] \\ \frac{-i}{k_2^2}\epsilon_{\mu}^b(k_g)\left ( \displaystyle\int \bar{u}(k)(ig\gamma_{\sigma}t^c)v(k') \psi_T(k,k^{\prime})dk\right )
\end{math} 
%2
\item
\begin{math} 
M_2 = \left ( \bar{v}(p_2) (ig\gamma_{\nu}t^a)\frac{i}{\slashed{k_1}-m}(ig\gamma_{\mu}t^b)\epsilon^{\mu,b}(k_g)u(p_1)\right )\\ (\frac{-ig^{\nu\sigma}}{k_2^2}) \left ( \displaystyle\int \bar{u}(k)(ig\gamma_{\sigma}{t^a})v(k')\psi_T(k,k')dk\right )
\end{math} 
%3
\item
\begin{math}
M_3 = \left ( \bar{v}(p_2)(ig\gamma_{\mu}t^b)\epsilon^{\mu,b}(k_g)\frac{i}{\slashed{k_1}-m}(ig\gamma_{\nu}t^a)u(p_1)\right )\\ (\frac{-ig^{\nu\sigma}}{k_2^2}) \left ( \displaystyle\int \bar{u}(k)(ig\gamma_{\sigma}{t^a})v(k')\psi_T(k,k')dk\right )
\end{math} 
%4
\item
\begin{math}
M_4 = \Big ( \bar{v}(p_2)(ig\gamma_{\nu}t^a)u(p_1)\Big ) (\frac{-ig^{\nu\sigma}}{k_2^2})\\ \Big ( \displaystyle\int \bar{u}(k)(ig\gamma_{\mu}t^b)\frac{i}{\slashed{k_1}-m}(ig\gamma_{\sigma}{t^a})v(k')\\ \psi_T(k,k')dk\Big ) \epsilon^{\mu,b}(k_g)
\end{math} 
%5
\item
\begin{math}
M_5 = \Big ( \bar{v}(p_2)(ig\gamma_{\nu}t^a)u(p_1) \Big ) (\frac{-ig^{\nu\sigma}}{k_2^2})\\ \Big ( \displaystyle\int \bar{u}(k)(ig\gamma_{\sigma}{t^a})\frac{i}{\slashed{k_1}-m}(ig\gamma_{\mu}t^b)v(k')\\ \psi_T(k,k')dk\Big ) \epsilon^{\mu,b}(k_g)
\end{math} 
%6
\item
\begin{math}
M_6 = \displaystyle\int \Bigg ( \left ( \bar{v}(p_2)(ig\gamma_{\nu}t^a)v(k') (\frac{-ig^{\nu\sigma}}{k_2^2}) \right ) \\ 
\left ( \bar{u}(k)(ig\gamma_{\sigma}{t^a})\frac{i}{\slashed{k_1}-m}(ig\gamma_{\mu}t^b)u(p_1)\right )\\ \psi_T(k,k') \Bigg ) dk \epsilon^{\mu,b}(k_g)
\end{math} 
%7
\item
\begin{math}
M_7 = \displaystyle\int \left ( \bar{v}(p_2)(ig\gamma_{\mu}t^b)\frac{i}{\slashed{k_1}-m}(ig\gamma_{\nu}t^a)v(k') \right )\\ (\frac{-ig^{\nu\sigma}}{k_2^2})  \Big ( \bar{u}(k)(ig\gamma_{\sigma}{t^a})u(p_1)\psi_T(k,k') \Big )dk\,\epsilon^{\mu,b}(k_g)
\end{math} 
%8
\item
\begin{math}
M_8 = \displaystyle\int \Bigg ( \bar{v}(p_2)(ig\gamma_{\nu}t^a)v(k') (\frac{-ig^{\nu\sigma}}{k_2^2})\\ \left ( \bar{u}(k)(ig\gamma_{\mu}t^b)\frac{i}{\slashed{k_1}-m}(ig\gamma_{\sigma}{t^a})u(p_1)\right ) \\ \psi_T(k,k')\Bigg ) dk \epsilon^{\mu,b}(k_g)
\end{math} 
%9
\item
\begin{math}
M_9 = \displaystyle\int \Bigg ( \left ( \bar{v}(p_2)(ig\gamma_{\nu}t^a)\frac{i}{\slashed{k_1}-m}(ig\gamma_{\mu}t^b)v(k')\right )\\ (\frac{-ig^{\nu\sigma}}{k_2^2}) \Big ( \bar{u}(k)(ig\gamma_{\sigma}{t^a})u(p_1)  \psi_T(k,k')\Big ) \Bigg ) dk\\ \epsilon^{\mu,b}(k_g)
\end{math} 
%10
\item
\begin{math}
M_{10} = \displaystyle\int \Bigg ( \Big ( \bar{v}(p_2)(ig\gamma_{\nu}t^a)v(k') gf^{bac} \Big ) \\ \Big [g^{\mu\nu}(-k_g - k_2)^{\sigma} + g^{\nu\sigma}(k_2 - k_1)^{\mu} + g^{\sigma\mu}(k_1 + k_g)^{\nu} \Big ]\\ (\frac{1}{k_2^2.k_1^2}) \Big ( \bar{u}(k)(ig\gamma_{\sigma}{t^a}u(p_1)\psi_T(k,k') \Big ) \Bigg )  dk\,\epsilon^{\mu,b}(k_g).
\end{math} 
\end{enumerate}
Here $g$ is the strong coupling constant.
%In the rest frame of $J/\psi$, the value of $k_2$ in diagrams 1 to 5 is $M_{J/\psi}$.
   The wavefunction $\psi_T(k,k')$ is obtained numerically, and hence analytical evaluation of the above expressions is not possible. The above amplitudes are evaluated explicitly. The averaged $M^2$ is then obtained as $\frac{1}{4\times9} \Sigma_{all~spins}\left [ \left (\Sigma_{i=1}^{10}M^*_i\right) \left (\Sigma_{i=1}^{10}M_i\right ) \right ]$. The factor $\frac{1}{4\times 9}$ comes from averaging over the spins and colors of the incoming charm quark and anti-charm quark. In order to avoid the ghost terms, only the transverse component of the gluon polarization term $\epsilon^{\mu,b}$ is used.

The charmonium wavefunction $\psi_T(k,k^{\prime})$ at temperature $T$ is obtained by solving the Schr\"{o}dinger equation with a phenomenological potential using the real part of the potential given in Eq.~\ref{eq:potential}. $m_c$ is the mass of charm quark, while $u$ and $v$ are the respective spinors of $c$ and $\bar{c}$ quarks, respectively.
%gans: rev2
The value of charm quark mass, $m_c$, used in the pQCD calculations as well as in other calculations (e.g.,  potential model computations) in this paper is 1.275 GeV.
%gans: rev2
 We note that $k=(\sqrt{|\vec{k}|^2 + m_c^2},\vec{k})$ and $k^{\prime} = (\sqrt{|\vec{k}|^2 + m_c^2},-\vec{k})$ as $4$-momentum of the charm and anti-charm quarks.  
%The above value of $M^2$ is multiplied by a factor of $\frac{1}{4\times 9}$ in order to average over the incoming quark and anti-quark spins and colors. 
%The factor $\frac{1}{4}$ in the expression for $M^2$ comes from averaging over the incoming quark and anti quark spins.
The recombination cross section in the laboratory frame is then computed in the cylindrical coordinates.
\begin{equation}
%\sigma(p_1,p_2).v_{rel} = \frac{1}{2\pi}\int_0^{2\pi} d\theta \int_0^{2\pi}d\beta \frac{1}{4\pi^2}\int \frac{1}{\gamma}d\vec{k}_{g\|}^{com} 
%\frac{M^2(p_1, p_2, k_g, k ).\vec{k}_{g\perp}}{2E_1.2E_2.E_{J/\psi}E_gh^{lab}}
\begin{split}
\sigma(p_1,p_2).v_{rel} = \frac{1}{2\pi}\int_0^{2\pi} d\theta \int_0^{2\pi}d\beta \frac{1}{4\pi^2}\int \Big [ \frac{1}{\gamma}d\vec{k}_{g\|}^{com} \times \\ 
\frac{M^2(p_1, p_2, k_g, k,k_{J/\psi} )\,\Theta(k^l_{J/\psi},k^h_{J/\psi})\,\vec{k}_{g\perp}}{2E_1\,2E_2\,E_{J/\psi}\,E_g\,h^{lab}} \Big ]
\end{split}
\end{equation}
\begin{equation}
\nonumber \Theta(k^l_{J/\psi},k^h_{J/\psi}) = \left \{
\begin{array}{l}
\nonumber   1~if~k^l_{J/\psi} < |\vec{k}_{J/\psi}| < k^h_{J/\psi}\\
%\nonumber \Theta(k^l_{J/\psi},k^h_{J/\psi}) = & 1~if~k^l_{J/\psi} < |\vec{k_{J/\psi}} < k^h_{J/\psi}\\
   0~otherwise.
\end{array}
\right .
\end{equation}
For ALICE suppression data at forward rapidity, we take $k^l_{J/\psi}$  to be $2.2\,$ GeV and $k^h_{J/\psi}$ to be a very large number, while at ALICE mid rapidity, the values are $0$ and $8$ GeV.  For the CMS suppression data, the values of $k^l_{J/\psi}$ and  $k^h_{J/\psi}$ are $6.5$ and $30\,$ GeV, respectively. 
The cylindrical coordinates are chosen so that $\vec{k}_{g\|}^{com}$ is along the direction of velocity of the center-of-mass ($\vec{v}^{com}$) of the $c\bar{c}$ system. The angle $\beta$ in the cylindrical coordinates is given by the angle of rotation around the $\vec{v}^{com}$ axis, and
$\gamma = \frac{1}{\sqrt(1 - |\vec{v}^{com}|^2)}$. $k_g,~k_{J/\psi},~p_1$ and $p_2$ are constrained by the $4$-momentum conserving delta function $\delta^4(p_1 + p_2 - k_g - k_{J/\psi})$. In general, for all the quantities, we use the notation $\vec{k}$ and $\vec{p}$ to denote the spatial component of the $4$-momentum $k$ and $p$.  
$h^{lab}$ is the outcome of integrating the energy conserving delta function $\delta(E_1 + E_2 - E_g - E_{J/\psi})$, and is given by 
\\
\begin{math}
h^{lab} = 1 + \frac{E_g}{(\gamma E_{J/\psi}^{com}.(1+\vec{v}^{com})^2)} - \frac{\vec{v}^{com}cos(\zeta)}{1+\vec{v}^{com}}
\end{math}
\\
$\zeta$ is the angle between $\vec{k}_{J/\psi}^{com}$ and $\vec{v}^{com}$ and $\theta$ is the angle between $c$ and $\bar{c}$ $3$-momenta in the laboratory frame.
The superscript "com" refers to the input $c\bar{c}$ center-of-mass frame.
%The infrared divergence due to the fermion propagator in the various Feynman diagrams cancels with each other. This is elaborated in the Appendix B.

We take the value of $\alpha = \frac{g^2}{4\pi} = 0.31$. 
%This  is primarily based on the expectation that the value of $\alpha$ would be around 0.3 near the midpoint of the ALICE momentum range, and secondly, based on fit to $J/\psi$ suppression data at LHC-ALICE.
%This value is the same value we use for the interaction of $J/\psi$ with the medium for collisional damping.  
The value of $\Gamma_{recomb}$ obtained as a function of temperature is shown in Fig. 2. The numerical values corresponding to Fig. 2 is given in Table I. 
\begin{table}
\caption {Temperature dependent values of $\Gamma_{recomb}$} 

\begin{tabular}{||c|c|c|c||}

\hline
Temperature  & $p_T < 8GeV$ & $p > 2.2GeV$ & $6.5 < p_T $  \\ 
(MeV) & (mb) & (mb)  &  (mb) \\ 
\hline
0   & 0.0000 & 0.0000 & 0.0000  \\ 
50  & 0.3624 & 0.0240 & 0.0000  \\ 
100 & 0.5697 & 0.1275 & 0.0003  \\ 
150 & 0.4823 & 0.1698 & 0.0017  \\ 
200 & 0.3706 & 0.1649 & 0.0040  \\
250 & 0.2378 & 0.1222 & 0.0052  \\
300 & 0.1219 & 0.0692 & 0.0044  \\
350 & 0.0389 & 0.0238 & 0.0020  \\
368 & 0.0224 & 0.0140 & 0.0013  \\ 
\hline

%old values
%\begin{tabular}{||c|c|c|c||}
%
%\hline
%Temperature  & $0 < p_T < 8GeV$ & $p > 2.2GeV$ & $6.5 < p_T < 30GeV$  \\ 
%(MeV) & (mb) & (mb)  &  (mb) \\ 
%\hline
%0   & 0.0000 & 0.0000 & 0.0000  \\ 
%50  & 0.4372 & 0.0289 & 0.0000  \\ 
%100 & 0.6873 & 0.1538 & 0.0003  \\ 
%150 & 0.5818 & 0.2049 & 0.0021  \\ 
%200 & 0.4471 & 0.1990 & 0.0048  \\
%250 & 0.2868 & 0.1474 & 0.0063  \\
%300 & 0.1470 & 0.0835 & 0.0053  \\
%350 & 0.0470 & 0.0287 & 0.0025  \\
%368 & 0.0270 & 0.0169 & 0.0016  \\ 
%\hline

\end{tabular}
\end{table}
\begin{figure}[h!]
%\vspace{2in}
%\includegraphics[width = 80mm,height = 80mm]{formation_time.eps}
\includegraphics[width = 80mm,height = 80mm]{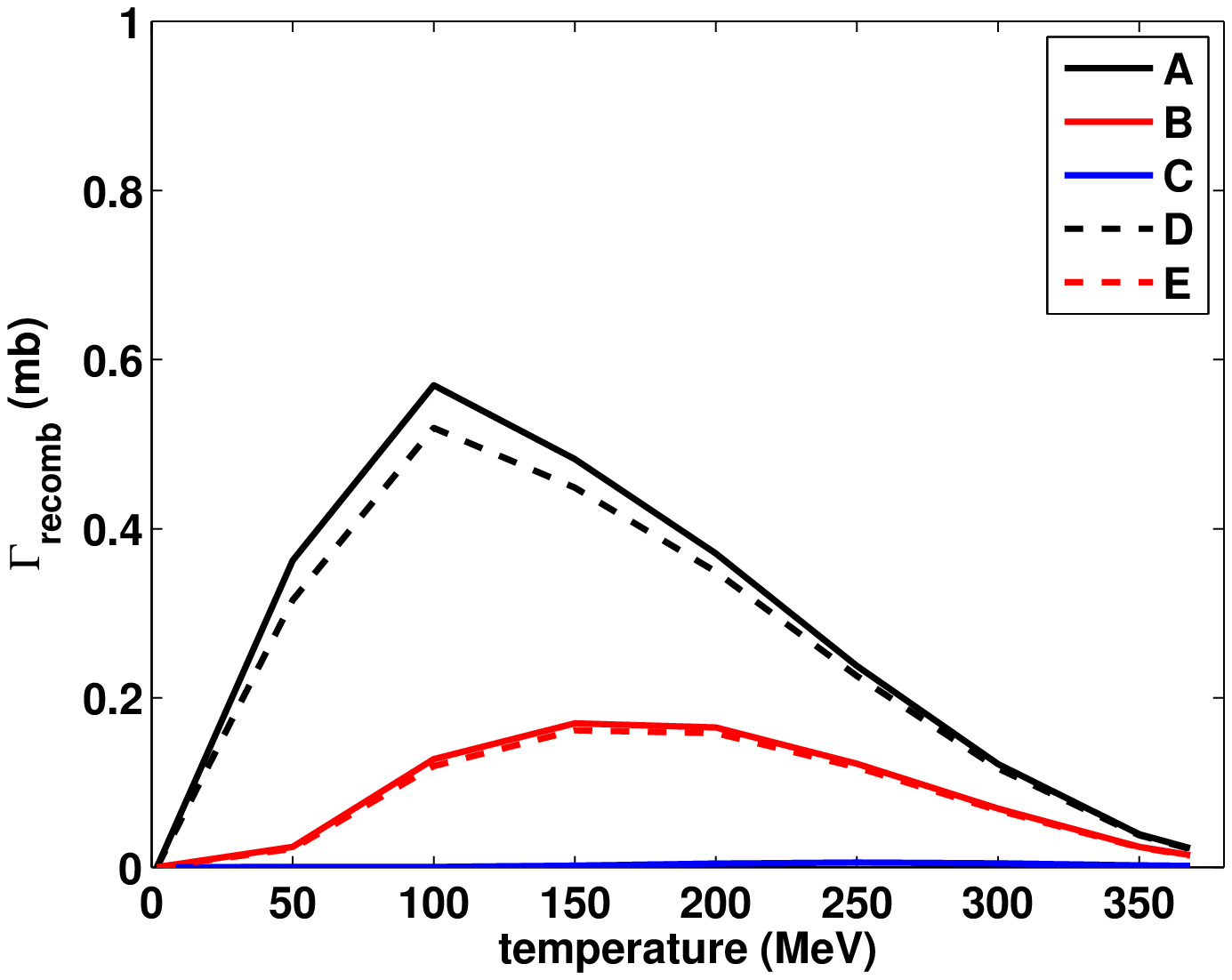}
\captionsetup{justification=raggedright, singlelinecheck=false}
\caption{$\Gamma_{recomb}$ as a function of temperature.\\The  legend is:\\
A = $\Gamma_{recomb}$ @ ALICE mid rapidity;\\
B = $\Gamma_{recomb}$ @ ALICE forward rapidity;\\
C = $\Gamma_{recomb}$ @ CMS;\\
D = $\Gamma_{recomb}$ @ ALICE mid rapidity, $\epsilon=1\,GeV$;\\
E = $\Gamma_{recomb}$ @ ALICE forward rapidity, $\epsilon=1\,GeV$ \\
$\epsilon$ = momentum cut off for virtual gluon propagator and outgoing gluon.}
\label{fig:formation_time}
\end{figure}
%The temperature averaged value of $\Gamma_{recomb}$ is about $0.53$ mb. 
%This value is comparable with $\Gamma_{recomb} = 0.65$ mb used in~\cite{comov}.
The value of $\Gamma_{recomb}$ at around $200$ MeV ($180$ MeV to $200$ MeV is about the lowest temperature of the QGP, and the recombination is likely to be highest at this temperature) is about $0.37$ mb at ALICE mid rapidity, $0.16$ mb at ALICE forward rapidity and $0.004$ mb at CMS. 
The range  $0.48$ mb to $0.37$ mb between the temperatures $150$ MeV to $200$ MeV, at ALICE mid rapidity is comparable with $\Gamma_{recomb} = 0.65$ mb used in~\cite{comov}, but on the lower side. The dashed lines indicate the value of $\Gamma_{recomb}$, when the amplitudes are zeroed out for gluon propagator virtuality or outgoing gluon momentum $< 1GeV$. The difference between the solid line and the corresponding dashed line is indicative of the number of soft gluons present.

Figure 3 shows the average value of $\sigma_{recomb}.v_{rel}$ for the special case of $|p_1| = |p_2|$ at $200$ MeV. 
Apart from the initial increase, the value of $\sigma_{recomb}$ decreases with the center-of-mass energy. 
The $\sigma_{recomb}$ for the ALICE mid rapidity is higher than the value for ALICE forward rapidity which is higher than the value for CMS mid rapidity. 

The decreasing value with energy is consistent with the fact that recombination decreases with increasing energy, as can be seen from the experimental data in Figs. 11, 12 and 13, where the observed recombination is highest at ALICE mid rapidity (lowest momentum range).
%The decreasing value with energy is consistent with the fact that the charmonium wavefunction mainly consists of low momentum charm quarks and anti-charm quarks, which had enabled us to model the charmonium wavefunction non-relativistically in the first place. The infrared divergence is handled by adding a small constant of  0.001 $GeV^2$ to the propagators in the denominator. 

%This points to a resonance when the sum of energy of charm and anti-charm quarks is equal to rest mass of $J/\psi$. 
\begin{figure}[h!]
%\vspace{2in}
%\includegraphics[width = 80mm,height = 80mm]{formation_time.eps}
\includegraphics[width = 80mm,height = 80mm]{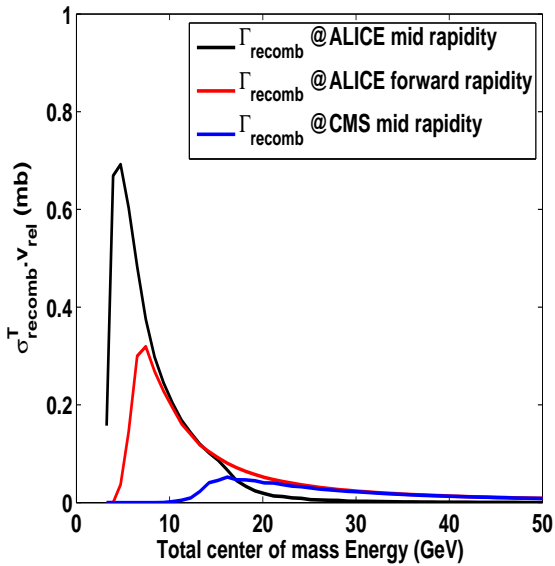}
\captionsetup{justification=raggedright, singlelinecheck=false}
\caption{$\Gamma_{recomb}$ as a function of total energy of c$\bar{c}$ at $200$ MeV}
\label{fig:formation_time}
\end{figure}
%gans  rev1 factorization
\subsubsection{Factorizability}
The formation of $J/\psi$ from ${c\bar{c}}$ pair is a non-perturbative effect involving exchange of soft gluons and occurs over a long time scale.  
Referring to Eq. B17 in Appendix B,  for soft outgoing gluons, the infrared cancellation is only partial. If ${\vec p}_1$ or ${\vec p}_2$ (the spatial components of $p_1$ and $p_2$ respectively) were zero, then the cancellation would have been exact. For the case of soft outgoing gluons, ${\vec p}_1$ or ${\vec p}_2$ is small but not zero. Hence, the cancellation of infrared divergence is partial. Consequently for soft gluons, factorization between the potential model based $J/\psi$ wavefunction and $c\bar{c}$  scattering would lead to  inaccuracy in the result.  
For very hard outgoing gluons, for the $c\bar{c}$ scattering sub-process, one can ignore the momentum spread of the $J/\psi$ wavefunction which is much smaller in comparison to the gluon momentum ($J/\psi$ momentum spread $\approx$ $0.5$ GeV at temperature of $170$ MeV). This means that the scattering of the $c$ and $\bar{c}$ occurs over a relatively short time scale. Hence, for very hard gluons, it is possible to factorize the potential model based $J/\psi$ wavefunction and the hard $c\bar{c}$ scattering process. At least, to the order $\alpha^3$ in the perturbation theory, it would then be possible to represent:\\
\begin{math}
\sigma(c\bar{c} \rightarrow J/\psi + g) \rightarrow \sigma(c\bar{c} \rightarrow c\bar{c} + g ) \psi^2(0),
\end{math}
\\
where $\psi(0)$ is the value of the $J/\psi$ wavefunction at $0$ spatial separation and captures the long duration interactions.
In the light of the above arguments, the validity of factorization of the recombination cross section would decrease in the order\\
CMS mid rapidity ($p_T >6.5$ GeV) $>$ ALICE forward rapidity ($p > 2.2$ GeV) $>$ ALICE  mid rapidity ($p_T > 0$ GeV).\\
In this work, in order to maximize the accuracy at ALICE conditions, where the recombination is high, factorization is not employed to determine the recombination cross section. 
%At higher orders of the perturbation expansion, a soft gluon propagator with one end connected to the outgoing gluon and the other end to the outgoing heavy quark (anti quark) line, can lead to IR divergences and subsequently breaking of the factorization even at the amplitude level. 
%page 31
% The authors in \cite{QQanni} argue that these divergences should cancel out. We believe that on similar grounds the IR divergences should cancel out for recombination also.

%gans end rev1 factorization

\subsection{Volume scaling computation}
The volume scaling is based on the quasi-particle model (QPM) equation of state (EOS) of the QGP and the concept of constant entropy condition. The QPM EOS used for computing the volume scaling is described in~\cite{Madhu2}.

In the QPM, the Reynolds number is given by\\
\begin{math}
R^{-1} = (4.0\,\eta)/(3.0\,T_0\,\tau_0\,s);
\end{math}
where $\eta = \frac{s}{4\pi}$, entropy density $s = 16.41$ GeV$^3$ and $T_0$ is the temperature at $\tau_0 = 0.5$ fm.
The volume is then given as:
\\
\begin{center}
\begin{math}
%V(\tau,b) = v_0( b).(\tau_0/\tau)^{R^{-1} - 1};
V(\tau) = v_0(\tau_0/\tau)^{R^{-1} - 1};
\end{math}
\end{center}
%where $v_0(b)$ is the initial volume at $\tau_0$, at impact parameter $b$, and is taken as\\ 
where $v_0$ is the initial volume at $\tau_0$,  and is taken as 
\begin{math}
%v_0(b) = \pi (R_{Pb} - b/2)^2 \tau_0,
%v_0(b) = \pi R_{Pb}^2 \tau_0,
v_0 = \pi R_{Pb}^2 \tau_0,
\end{math}
with $R_{Pb}$ = $6.62$ fm from the Woods-Saxon distribution~\cite{woods}.
%In actual heavy-ion experiments, the volume, the thermalization time and the lifetime of the QGP would depend on the centrality bin. However, as an approximation, we take these quantities to be independent of the centrality bins since their precise centrality dependence is still not known.

\subsection{Lighter quarks and gluons}
\label{sec:light}
The lighter quarks and gluons would also form $J/\psi$ and $c\bar{c}$. After including these processes, the rate equations get modified as
\begin{eqnarray} 
\nonumber \frac{dN_{J/\psi}}{dt} = -\Gamma_{diss} N_{J/\psi}(t) + \Gamma_{recomb} \frac{N_c(t) N_{\bar{c}}(t)}{V(t)} +\\ \nonumber
\nonumber\Lambda_{q,g\rightarrow J/\psi},\\ \nonumber
\nonumber\frac{dN_c}{dt} = \Gamma_{diss} N_{J/\psi}(t) - \Gamma_{recomb} \frac{N_c(t) N_{\bar{c}}(t)}{V(t)} +\\ \Lambda_{q,g\rightarrow c\bar{c}},\nonumber
\end{eqnarray} 
where
\begin{description}
\item
\begin{eqnarray}
%	\Lambda_{q,g\rightarrow J/\psi} = \int dp_1 dp_2 f_p(p_1)f_p(p_2)\sigma^T_{q,g\rightarrow J/\psi}(p_1,p_2) \\ 
%~~~~~~~~~~~~~~~~v_{rel}^{pp}(p_1,p_2)
\nonumber	\Lambda_{q,g\rightarrow J/\psi} = \int dp_1 dp_2 \Big [ f_p(p_1)f_p(p_2)
 \times \\ \nonumber \sigma^T_{q,g\rightarrow J/\psi}(p_1,p_2) v_{rel}^{pp}(p_1,p_2) \Big ] 
\end{eqnarray}
\item
\begin{eqnarray}
%	\Lambda_{q,g\rightarrow c\bar{c}} = \int dp_1 dp_2 f_p(p_1)f_p(p_2)\sigma^T_{q,g\rightarrow c\bar{c}}(p_1,p_2) \\ ~~~~~~~~~~~~~~~~v_{rel}^{pp}(p_1,p_2)
\nonumber	\Lambda_{q,g\rightarrow c\bar{c}} = \int dp_1 dp_2 \Big [ f_p(p_1)f_p(p_2) \times \\ \nonumber \sigma^T_{q,g\rightarrow c\bar{c}}(p_1,p_2) v_{rel}^{pp}(p_1,p_2) \Big ] 
\end{eqnarray}
\end{description}
The light quarks and gluons are assumed to be thermalized and hence momentum distribution $f_p(p_1)$ and $f_p(p_2)$ are taken to be Fermi Dirac distribution$\frac{g_q}{exp( (E - \mu)/T) + 1}$ for light quarks and anti-quarks and Bose Einstein $\frac{g_g}{exp( (E - \mu)/T) - 1}$ for the gluons. The  baryonic chemical potential $\mu$ is taken to be vanishingly small at the LHC. The value of degeneracy factor for light quarks ($g_q$) and for gluons ($g_g$) is taken as $6$ and $16$, respectively.

For the lighter quarks, only the first five of the ten Feynman diagrams are applicable. With the first five diagrams, the value of $\sigma^T_{q,g\rightarrow J/\psi}(p_1,p_2)$ for the lighter quarks are not expected to be negligible when compared to that of $c\bar{c} \rightarrow J/\psi$ cross section. 

However, with vanishing value of $\mu$, it is seen that most of the quarks and gluons lie in the low energy region. The semi-log plot in Fig. 4 shows that at around $1.5$ GeV, the particle density reduces to around $10^{-5}$ GeV$^{-4}$.
\begin{figure}[h!]
%\vspace{2in}
%\includegraphics[width = 80mm,height = 80mm]{formation_time.eps}
\includegraphics[width = 80mm,height = 80mm]{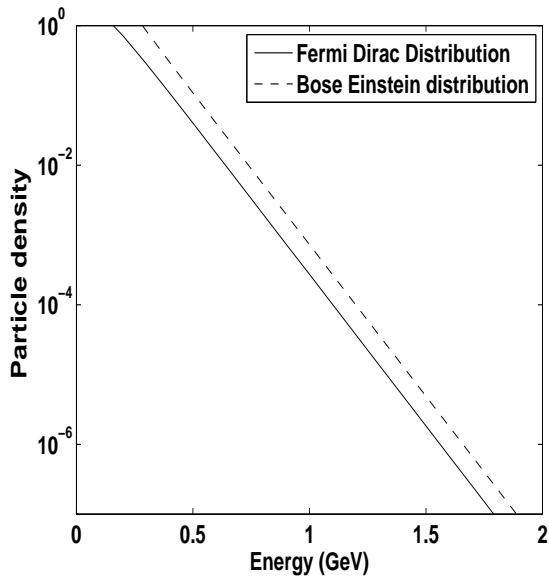}
\captionsetup{justification=raggedright, singlelinecheck=false}
\caption{Fermi Dirac and Bose Einstein distribution }
\label{fig:fermi_bose}
\end{figure}
This is not sufficient to form either $J/\psi$ or $c\bar{c}$ in any significant quantity, resulting in a negligible value for $\Lambda_{q,g\rightarrow J/\psi}$ and $\Lambda_{q,g\rightarrow c\bar{c}}$. 
%The values of  $\Lambda_{q,g\rightarrow J/\psi}$ and $\Lambda_{q,g\rightarrow c\bar{c}}$ are found to be about 1000 times smaller than $\Lambda_{recomb}$ for the lighter quarks.
  In view of the above facts, with a Fermi-Dirac or Bose-Einstein distribution for a thermalized quarks and gluon, we neglect the contribution arising due to the light quarks and gluons to the rate equation. 

\section{Gluonic-dissociation, Collisional damping, CNM and Color Screening}
\subsection{Gluonic-dissociation, Collisional damping and CNM effect}
The modeling of Gluonic-dissociation, collisional damping are on the same lines as~\cite{rituraj}. For ease of reference, we reproduce the quark, antiquark singlet potential used in this work~\cite{Wolschin}. 
\begin{equation}
\label{eq:potential}
\begin{split}
	V(r,m_D) = \frac{\sigma_{string}}{m_D}(1 - e^{-m_D\,r}) -\\
\alpha_{eff} \left ( m_D + \frac{e^{-m_D\,r}}{r} \right ) - \\
i\alpha_{eff} T \int_0^\infty \frac{2\,z\,dz}{(1+z^2)^2} 
\left ( 1 - \frac{\sin(m_D\,r\,z)}{m_D\,r\,z} \right ),
\end{split}
\end{equation} 
where $m_D$ is the Debye mass given by 
\begin{math}
    m_D = T\sqrt{4\pi \alpha_s^T \left ( \frac{N_c}{3} + \frac{N_f}{6} \right ) }.
\end{math}
$N_f = 3$ = number of flavors; $\alpha_s^T = 0.49$; $\sigma_{string}= 0.192$ GeV$^2$.
$\alpha_{eff} = \frac{4\alpha}{3}$, where we have taken $\alpha = 0.22$. This value of alpha, along with $\alpha_s^T = 0.49$, gives the dissociation temperatures close to the values $381$ MeV for $J/\psi$, $190$ MeV for $\psi^{\prime}$, and $197$ MeV for $\chi_c$~\cite{Madhu1}.

The CNM effect is also on the same lines as~\cite{rituraj}.
The CNM formulation in the current work and in Ref.~\cite{rituraj} is based on the work done by~\cite{vogt,newvogt}. 
The rapidity distribution for the computation of $x_1$ and $x_2$  is taken to be $|y| < 0.9$ for ALICE mid rapidity, $2.5 < |y| < 4$ for ALICE forward rapidity and $|y| < 2.4$ for CMS. 
The shadowing parametrization EPS09~\cite{EPS09} is used to obtain the shadowing function, and the The gluon distribution function in a proton  has been estimated using CTEQ6~\cite{CTEQ6}.

\subsection{Color Screening}
\label{sec:rev1:color}
%gans rev1
The color screening model in this work differs from ~\cite{rituraj} in terms of the use of temperature dependent formation time.
%gans rev1
The color screening model used in the present work and in Ref.~\cite{rituraj} is based on pressure profile~\cite{Madhu1} in the transverse plane and cooling law for pressure based on QPM EOS~\cite{gans,Madhu2} for QGP. The cooling law for pressure is given by
\begin{equation}
	p(\tau,r) = A + \frac{B}{\tau^q} + \frac{C}{\tau} + \frac{D}{\tau^{c_s^2}},
\end{equation}
where $A = -c_1$, $B = c_2\,c_s^2$, $C = \frac{4\eta q}{3(c_s^2 - 1)}$ and $D = c_3$. The constants $c_1$, $c_2$, $c_3$ are given by
\begin{itemize}
\item $c_1 = - c_2\tau'^{-q} - \frac{4\eta}{3c_s^2\tau'}$; 
\item $c_2 = \frac{\epsilon_0 - \frac{4\eta}{3c_s^2}\left( \frac{1}{\tau_0} - \frac{1}{\tau'} \right ) }{\tau_0^{-q} - \tau'^{-q}}$
; 
\item $c_3 = \left (p_0 + c_1 \right) \tau_0^{c_s^2} - c_2\,c_s^2 \tau_0^{-1} - \frac{4\eta}{3}\left ( \frac{q}{c_s^2 - 1} \right ) \tau_i^{\left ( c_s^2 - 1\right)}$.
%\item $c_3 = \left (p_0 + c_1 \right) \tau_0^{c_s^2} - c_2c_s^2 \tau_0^{-1} - \frac{4\eta}{3}\left ( \frac{q}{c_s^2 - 1} \right ) \tau_0^{\left ( c_s^2 - 1\right )}$.
\end{itemize}
The above constants are determined by using different boundary conditions on pressure and energy density described in~\cite{gans,Madhu2}. 

Writing the above equations at initial time $\tau = \tau_0$ and screening time $\tau = \tau_s$ and combining with the pressure profile~\cite{Madhu2}, we get the following two equations:

%\begin{equation}
\begin{eqnarray}
	p(\tau_0,r) = A + \frac{B}{\tau_0^q} + \frac{C}{\tau_0} + \frac{D}{\tau_0^{c_s^2}} = p(\tau_0,0)\,h(r);
%\end{equation}
\end{eqnarray}
\begin{eqnarray}
	p(\tau_s,r) = A + \frac{B}{\tau_s^q} + \frac{C}{\tau_s} + \frac{D}{\tau_s^{c_s^2}} = p_{QGP},
\end{eqnarray}
where $p_{QGP}$ is the pressure of QGP inside the screening region required to dissociate a particular charmonium state and it is determined by QPM EOS for QGP medium. The above equations are solved numerically and the screening time $t_s$ is equated with the charmonium formation time $t_f$ = $\gamma \tau_f(T)$ to determine radius of screening region $r_s$~\cite{Madhu2,gans, gans2}.
$\gamma$ is the Lorentz factor corresponding to the transverse energy of $J/\psi$.
The temperature dependent formation time $t_F(T)$ is determined using the simulation of time independent Schr\"{o}dinger equation as~\cite{gans2}
\begin{equation}
\tau_f(T_1) = \tau_f(T_0)\frac{E_{bind}(T_0)}{E_{bind}(T_1)}.
\end{equation}
\begin{figure}[h!]
\includegraphics[width = 80mm,height = 80mm]{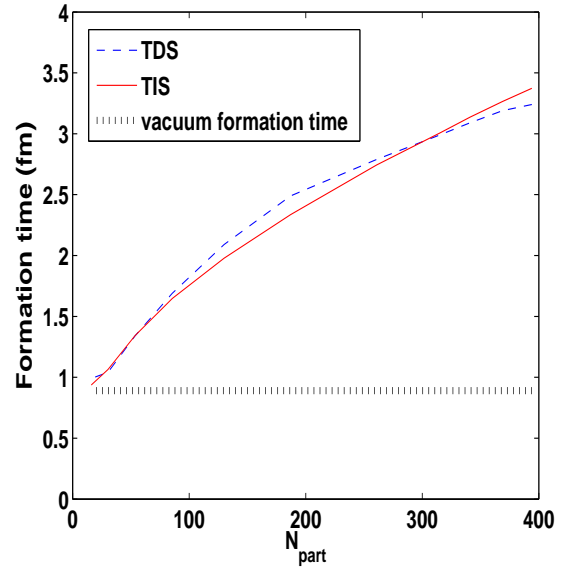}
\captionsetup{justification=raggedright, singlelinecheck=false}
\caption{Comparison of the $J/\psi$ formation time between the time dependent Schr\"{o}dinger equation (TDS), the time independent Schr\"{o}dinger equation (TIS) and the vacuum formation time.}
\label{fig:form_time}
\end{figure}
Figure 5 depicts the formation time for $J/\psi$ obtained from both the time independent Schr\"{o}dinger equation and the time dependent Schr\"{o}dinger equation. The temperature associated with the initial conditions, $T_{init}$~\cite{gans2}, is taken as a value between $200$ MeV and $150$ MeV based on the centrality.
The expression for survival probability due to color screening can be obtained as: 
\begin{eqnarray}
\nonumber    S_c(p_T,N_{part}) = \frac{2(\alpha + 1)}{\pi R_T^2} 
\int_0^{R_T} dr\, r \,\phi_{max}(r) \\ \nonumber 
\left \{ 1 - \frac{r^2}{R_T^2} \right \}^\alpha, 
\end{eqnarray}
where $\alpha=0.5$, $R_T$ and $\phi_{max}$ (which is a function of $p_t$ and $r_s$) are defined in~\cite{gans,Madhu2}. 
%The final suppression due to color screening for $J/\psi$ is obtained after including the feeddown contribution from higher resonance states of charmonium.

\subsection{Final suppression}
In the computation of the final suppression, all the above mentioned effects are included.

In Fig. 2, it is shown that the recombination rate $\Gamma_{recomb}$ decreases with temperature at high temperature range, again as a consequence of Debye mass $m_D$ widening (spatially) the $J/\psi$ waveform. This raises the question whether the decreased recombination rate is remodeling the suppression due to color screening mechanism described in section III (D).
The temperature dependent color screened recombination cross section is based on the interaction of uncorrelated charm quark and anti-charm quark produced in the latter stages of the QGP, and thus is unrelated to the color screening mechanism (as described in section III (D) ), which acts upon the correlated charm and anti-charm quark pair produced during the initial hard collisions. This implies that the processes of color screening and recombination apply to the different non-overlapping population of charm quarks (correlated and uncorrelated). Hence, both mechanisms are required to model the charmonium suppression in the QGP.
%Furthermore, if the temporal regions happen to be non-overlapping, it further helps in ensuring that the same physics is not modeled by both color screening and temperature dependent recombination cross section.

The effect of shadowing and the rate equation involving gluonic dissociation, collisional damping and recombination, can be put together as:
\begin{math} 
S_{cnm}S_{dr}.
\end{math}
The question arises as to how the suppression due to the color screening can be combined. Given the large temperature dependent formation time, the process of gluonic dissociation and collisional damping can happen even before $J/\psi$ is fully formed. Hence, the processes of color screening and gluonic dissociation and collisional damping may overlap in time. 

In order to incorporate the color screening into the system of rate equations within the framework of the above arguments, we first determine the equivalent $\Gamma_{c}$ corresponding to $S_c$. 
\begin{center}
\begin{math}
%\Gamma_c = -\ln(S_c)/(\gamma \tau_{f} - t_0).
\Gamma_c = -\ln(S_c)/(t_{QGP} - t_0).
\end{math}
\end{center}
We then have $\Gamma_{diss} = \Gamma_{dg} + \Gamma_c$,
%gans rev2
where $\Gamma_{dg} = \Gamma_{diss,nl} + \Gamma_{damp}$, with $\Gamma_{diss,nl}$ and $\Gamma_{damp}$ being gluodissociation and collisional damping dissociation rates~\cite{rituraj,gans2}.  
%gans rev2
At ALICE forward rapidity conditions, due to the non-applicability of Bjorken's scaling law at forward rapidity, we ignore the color screening mechanism. 

%ALICEjpsi
\section{Results and Discussions}
\label{sec:results}
The experimental charm quark distribution in Fig. 1 is a static distribution, and does not vary with temperature or time. To overcome this difficulty, the modified Fermi Dirac distribution $N(\lambda)/(exp(\lambda E/kT) + 1)$ is used, where $N(\lambda)$ is a normalization factor. The value of $\lambda$ is fixed by comparing the distribution to the experimentally determined distribution at temperature $T_c = 180$ MeV. The modified Fermi Dirac distribution is employed in all the simulations. 
The Tsallis distribution,
\begin{math}
N(\lambda,q) \left [1 + (1-q)(-\lambda E/T) \right ] ^{\frac{1}{1-q}},
\end{math}
with $q \rightarrow 1+$, gives very similar results (not shown in this paper).
Our calculation of the temperature dependent recombination cross-section is shown in Fig. 2. As the temperature approaches the dissociation temperature of $J/\psi$, the value of $\Gamma_{recomb}$ approaches zero in accordance with the expectation. 
At very low temperature, near $0$ MeV, the charm quark and anti-quark do not have sufficient energy to form $J/\psi$, and hence, the $\Gamma_{recomb}$ approaches $0$ at low temperature also. 
Qualitatively, the temperature behavior is similar to the temperature dependence obtained in Ref.~\cite{tempform} for $b\bar{c},\bar{b}c$ bound states. 
%We do however suspect that at low temperature close to 0MeV, non-perturbative QCD may play a significant role due to low momentum. 
The relevant temperature for QGP would be above $T_c = 180 MeV$.
It is also seen from Fig. 3, that at high momentum range at the CMS, the cross section is very small compared to the cross section at the lower momentum range of the ALICE experiments at both forward and mid rapidity. The cross section is highest at the ALICE mid rapidity experiment, which has the lowest momentum range.  

The predicted $J/\psi$ suppression due to color screening alone is shown in Figs. 6 and 7. The experimental data on $J/\psi$ suppression for ALICE mid rapidity~\cite{ALICEjpsi} and CMS~\cite{PKSCMS1} at the LHC energy $\sqrt{s_{NN}}=2.76$ TeV versus $N_{part}$ is also shown for comparison. Similar to the case of $\Upsilon$, with temperature dependent formation time~\cite{gans2}, the effect of suppression due to color screening mechanism becomes very small or almost disappears. This result forms the basis of ignoring color screening at ALICE forward rapidity. 
Color screening with either temperature dependent formation time or constant formation time does not capture the suppression for both ALICE and CMS data simultaneously. This indicates some other processes may be playing a role. 
%\begin{figure}[h!]
%\includegraphics[width = 80mm,height = 80mm]{only_colour_alice.eps}
%\captionsetup{justification=raggedright, singlelinecheck=false}
%\caption{Our predicted suppression due to only color screening. Experimental data on $J/\psi$ suppression at LHC are taken from~\cite{ALICEjpsi}.}
%\label{fig:color_alice}
%\end{figure}
\begin{figure}[h!]
%\vspace{2in}
%\includegraphics[width = 80mm,height = 80mm]{formation_time.eps}
\includegraphics[width = 80mm,height = 80mm]{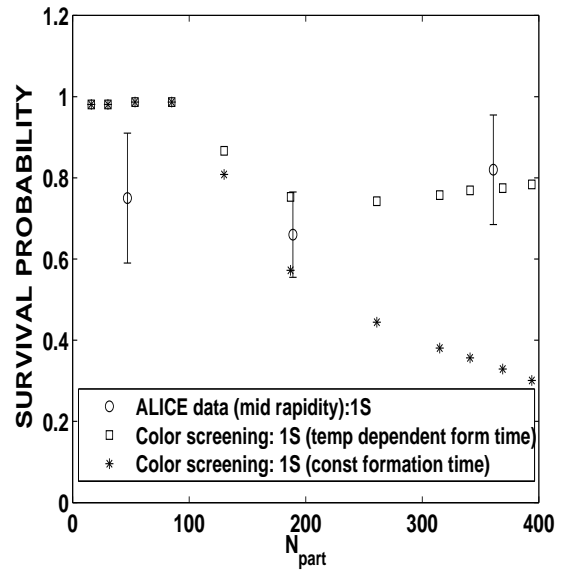}
\captionsetup{justification=raggedright, singlelinecheck=false}
\caption{Our predicted suppression due to only color screening. Experimental data on $J/\psi$ suppression at ALICE (mid rapidity) LHC are taken from~\cite{ALICEjpsi}.}
\label{fig:color}
\end{figure}
\begin{figure}[h!]
%\vspace{2in}
%\includegraphics[width = 80mm,height = 80mm]{formation_time.eps}
\includegraphics[width = 80mm,height = 80mm]{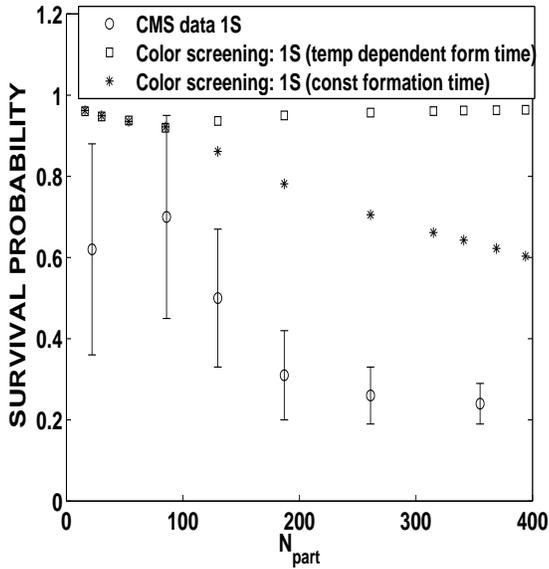}
\captionsetup{justification=raggedright, singlelinecheck=false}
\caption{Our predicted suppression due to only color screening. Experimental data on $J/\psi$ suppression at CMS LHC are taken from~\cite{PKSCMS1}.}
\label{fig:color}
\end{figure}

Figure 8 shows the effect of shadowing, dissociation and recombination for the ALICE mid rapidity (low momentum range). The curve "only dissociation" is obtained by assigning $\Gamma_{recomb} = 0$. The difference between the "only dissociation" and "recombination and dissociation" curves is the result of recombination.
\begin{figure}[h!]
%\vspace{2in}
%\includegraphics[width = 80mm,height = 80mm]{formation_time.eps}
\includegraphics[width = 80mm,height = 80mm]{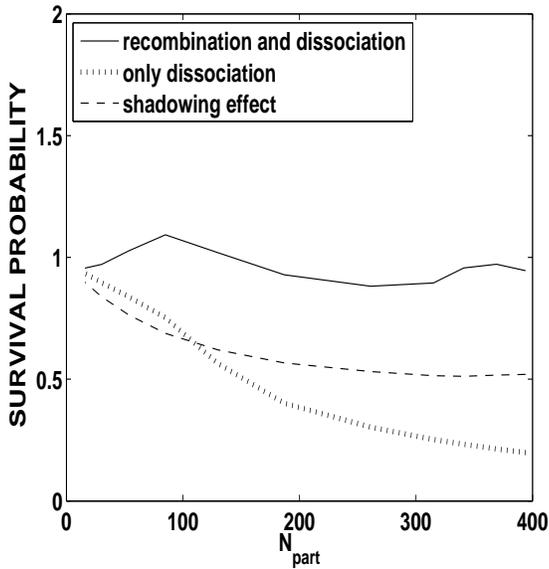}
\captionsetup{justification=raggedright, singlelinecheck=false}
\caption{Comparison between the effect of color screening, recombination and shadowing at momentum values corresponding to ALICE data at mid rapidity.}
\label{fig:individual_alice}
\end{figure}
\begin{figure}[h!]
%\vspace{2in}
%\includegraphics[width = 80mm,height = 80mm]{formation_time.eps}
\includegraphics[width = 80mm,height = 80mm]{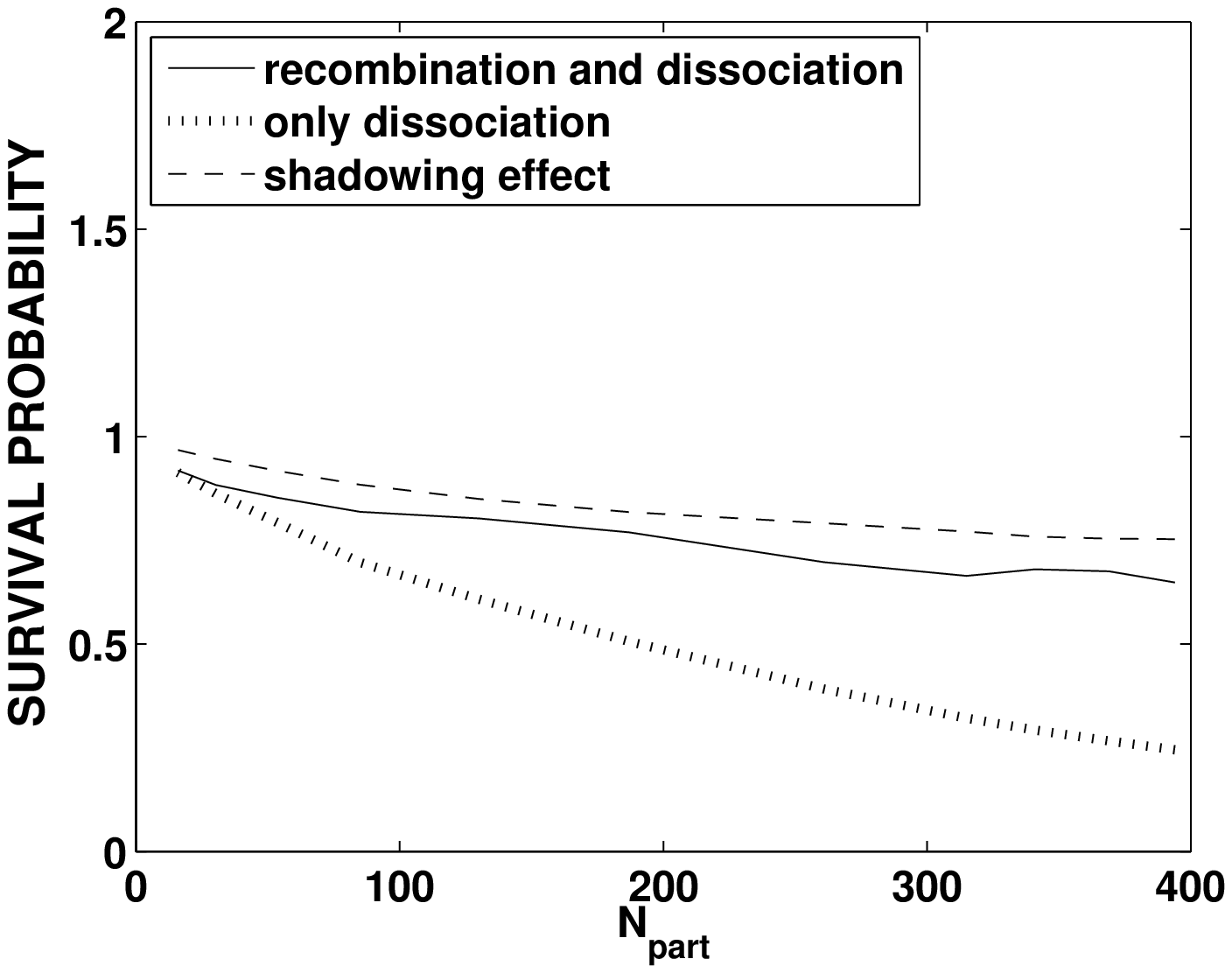}
\captionsetup{justification=raggedright, singlelinecheck=false}
\caption{Comparison between the effect of color screening, recombination and shadowing at momentum values corresponding to ALICE data at forward rapidity.}
\label{fig:individual_alice}
\end{figure}

\begin{figure}[h!]
\includegraphics[width = 80mm,height = 80mm]{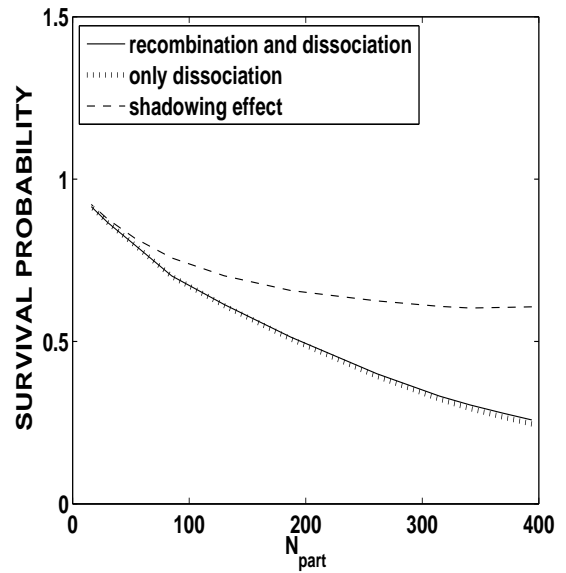}
\captionsetup{justification=raggedright, singlelinecheck=false}
\caption{Comparison between the effect of color screening, recombination and shadowing at momentum values corresponding to CMS data. 
The curves "recombination and dissociation" and "only dissociation" almost overlap, indicating negligible recombination.}
\label{fig:cms_individual}
\end{figure}
It can be seen that the recombination increases with centrality. 
Recombination depends on the product $N_c\,N_{\bar{c}}$. If we set $N_c = N_{\bar{c}}$, recombination depends quadratically on $N_c$. On the other hand dissociation depends linearly on $N_{J/\psi}$. In the peripheral collisions both $N_{J/\psi}$ and $N_c$ are small, but due to the quadratic dependence on $N_c$ recombination is lower in the peripheral collisions. 
Figures 9 and 10 show the effect of shadowing, dissociation and recombination at ALICE at forward rapidity and at CMS at mid rapidity. Recombination is seen to be smaller at ALICE forward rapidity than at ALICE  mid rapidity, and becomes negligible at the CMS where the momentum range is very high. The dependence of recombination with centrality at ALICE forward rapidity remains similar to ALICE mid rapidity. 

We now compare the final suppression data of $J/\psi$ at ALICE mid and forward rapidity in Figs. 11 and 12.
In the central region where recombination is highest, it can be seen that $J/\psi$ suppression at forward rapidity is in reasonable agreement with experimental data, but at ALICE mid rapidity, the suppression is overestimated due to the recombination being underestimated. 
As discussed earlier, the longitudinal component of momentum gives rise to higher momentum at forward rapidity as compared to mid rapidity for the same $p_T$. Due to the larger momentum, pQCD calculations at forward rapidity are more reliable than at mid rapidity. Clearly non-perturbative QCD is required to explain the recombination data more accurately at ALICE mid-rapidity. 

Finally, Fig. 13 shows the final suppression of $J/\psi$, which is comparable to the experimental CMS data~\cite{PKSCMS1} up to a large extent throughout the whole range of $N_{part}$. Recombination was seen to be almost vanishing for the CMS momentum range in Fig. 10.

In summary, we have been able to explain to a reasonable good extent, the $J/\psi$ suppression at ALICE forward rapidity and CMS mid rapidity. Even though the $J/\psi$ suppression data has not been explained at ALICE mid rapidity accurately, it can be inferred from Figs. 8, 9 and 10, that pQCD correctly predicts the decreasing trend in recombination from ALICE mid rapidity to ALICE forward rapidity to CMS mid rapidity.   
\begin{figure}[h!]
%\vspace{2in}
%\includegraphics[width = 80mm,height = 80mm]{formation_time.eps}
\includegraphics[width = 80mm,height = 80mm]{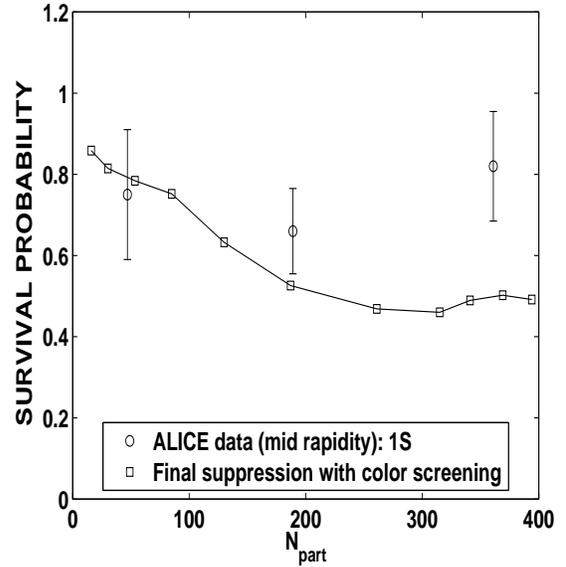}
\captionsetup{justification=raggedright, singlelinecheck=false}
\caption{Final suppression with color screening, gluonic dissociation along with collisional damping, recombination and shadowing. Experimental data on suppression at ALICE at mid rapidity are taken from~\cite{ALICEjpsi}.}
\label{fig:all_alice}
\end{figure}
\begin{figure}[h!]
%\vspace{2in}
%\includegraphics[width = 80mm,height = 80mm]{formation_time.eps}
\includegraphics[width = 80mm,height = 80mm]{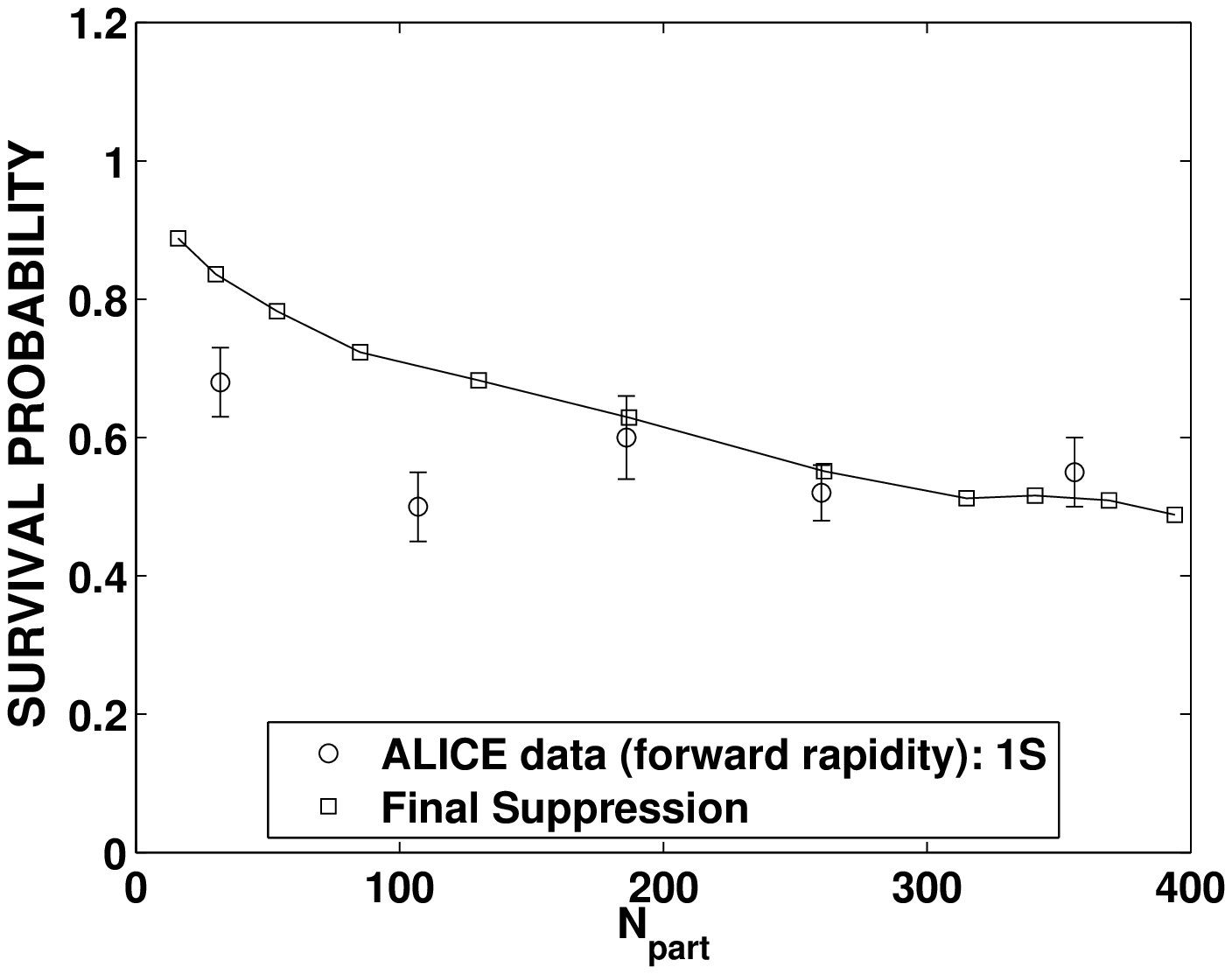}
\captionsetup{justification=raggedright, singlelinecheck=false}
\caption{Final suppression with color screening, gluonic dissociation along with collisional damping, recombination and shadowing. Experimental data on suppression at ALICE at forward rapidity are taken from~\cite{ALICEfor}.}
\label{fig:all_alice_for}
\end{figure}
\begin{figure}[h!]
%\vspace{2in}
%\includegraphics[width = 80mm,height = 80mm]{formation_time.eps}
\includegraphics[width = 80mm,height = 80mm]{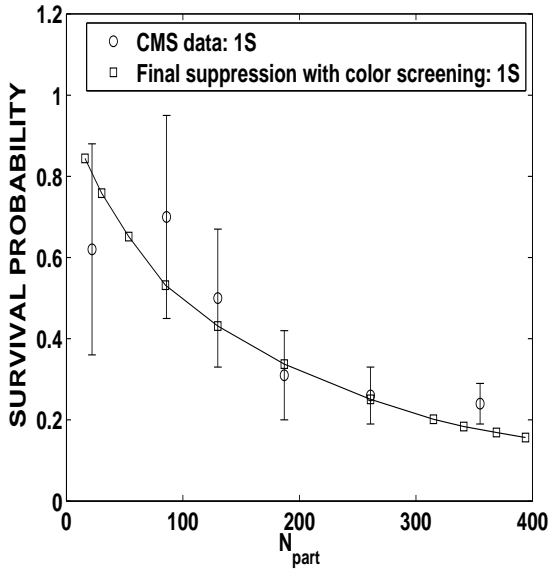}
\captionsetup{justification=raggedright, singlelinecheck=false}
\caption{Final suppression with color screening, gluonic dissociation along with collisional damping, recombination and shadowing. Experimental data on suppression at CMS are taken from~\cite{PKSCMS1}.}
\label{fig:cms_all}
\end{figure}

%gans rev1 result_comp
Comparison of Fig.~\ref{fig:cms_all} with Fig. 10 in Ref.~\cite{rituraj} shows similar suppression of $J/\psi$. However, the figure in Ref.~\cite{rituraj} indicates that the recombination is more than 20\% for certain centrality bins. 
The recombination cross section based on pQCD calculation results in only about 2\% increase in relative $J/\psi$ yield due to recombination at CMS mid rapidity (Fig.~\ref{fig:cms_individual}). Experimental data relating $J/\psi$ $R_{AA}$ to $p_T$ at the CMS ~\cite{ptrelate} indicate recombination may become very small at high $p_T$.
The increase in suppression due to lesser recombination is offset by a decrease in suppression due to color screening due based on temperature dependent formation time. 
Coincidentally, the net $J/\psi$ suppression at CMS mid rapidity is similar between the current work and~\cite{rituraj}.
%gans rev1 result_comp

Apart from the discrepancy due to pQCD approximation, an additional source of uncertainty in suppression might be that strange quarks may not be fully thermalized. If we compare the non-thermalized distribution in Fig. 1 and the thermalized Fermi Dirac or Bose Einstein distribution in Fig. 4, we can see that the distribution is shifted to the right in case the distribution is non-thermalized. A partially non-thermalized strange quark distribution will have more strange quarks with sufficient energy to form $c\bar{c}$ pairs or $J/\psi$. This would lead to a non-negligible value of $\Lambda_{q,g\rightarrow c\bar{c}}$ and $\Lambda_{q,g\rightarrow J/\psi}$. 
This would infuse more $c\bar{c}$ and $J/\psi$ into the system, leading to a slightly different value of suppression. 
%If we look at Fig. 1, we find that the charm quark distribution peaks at the energy of around $2.3$ GeV to $3.3$ GeV. This corresponds to a total energy of $4.6$ GeV to $6.6$ GeV for the $c\bar{c}$ pair. 
Comparing the experimental charm quark distribution and the modified Fermi Dirac distribution, there is a substantial difference in the population of charm quark at lower energy. Since the momentum range at ALICE forward rapidity and CMS mid rapidity is on the higher side, this  difference should not impact the recombination calculation considerably for these two cases. The impact on ALICE mid rapidity could however be higher. 
It is also observed that shadowing calculation is a relatively uncertain quantity. Though we have not depicted the uncertainty in this work, {\bf EPS09} parametrization gives a range of uncertainty in the shadowing. 
%Again, with increase in energy, the cross section increases, which implies that in higher energy collisions than 2.76TeV, one might see a higher amount of recombination.
%One reason could be the overestimation of the initial number of $c\bar{c}$ pairs. We currently estimate $c\bar{c}$ pairs from D meson measurements. 
%However these D meson numbers are not the numbers at the initial time of the p-p collisions,  but rather is the cumulative number obtained at the end of the p-p collision. Some number of $c\bar{c}$ would have formed after the initial time of the p-p collision. Since the rate equations depend quadratically on the number $N_c$, an overestimation of recombination could be expected.   

%gans rev1 hadronization 
Another source of uncertainty may be the formation of $J/\psi$ during hadronization phase. To analyze this uncertainty, let us look at the formation of open charm mesons during the QGP phase and hadronization phase.
The dissociation time of open charm mesons at the QGP deconfinement temperatures is expected to be very small (less than a fraction of a fermi) compared to that of hidden charm mesons, and thus any open charm meson in the QGP medium will disappear quickly~\cite{thews}. 
%Kinetic formation model; page 23
One may safely assume that  open charm mesons would be formed mainly during hadronization. During hadronization, due to the huge abundance of light quarks (possibly 2 or 3 orders of magnitude more than charm quarks), formation of open charm meson would be predominantly much more than the formation of $J/\psi$ or any other hidden charm mesons. 
This can be seen from comparison of $J/\psi$ and $D$ meson production from pp collision experiments too~\cite{PKSCMS1,ccbar}, after adjusting for luminosity. 
As a result, the number of charm quark or anti-quark (=$N_c$) available for recombination to form $J/\psi$ would reduce substantially. If one were to continue with the quadratic dependence on available $N_c$ for recombination during hadronization, then the number of $J/\psi$ formed due to recombination during hadronization would be negligible compared to the number of $J/\psi$ formed due to recombination during the QGP phase. 
In this proposed model, we thus ignore the formation of $J/\psi$ during hadronization, which may lead to some discrepancy.

Despite the above uncertainties, we see that our model has been able to explain the trend of centrality dependent $J/\psi$ suppression data in the forward rapidity region obtained from the ALICE experiment and mid rapidity region from the CMS experiments with reasonable agreement. 
%gans rev1 hadronization 

\begin{figure}[h!]
\includegraphics[width = 80mm,height = 80mm]{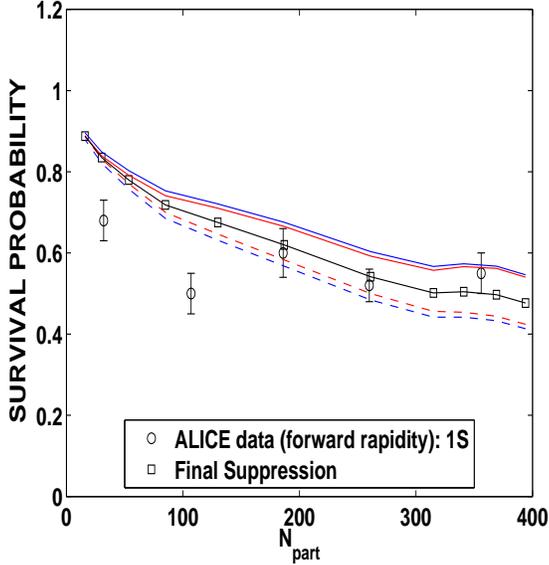}
\captionsetup{justification=raggedright, singlelinecheck=false}
\caption{Error propagation: The red curves indicate 10\% change in initial conditions, while the blue curves indicate 10\% change in dissociation/recombination constants. Detailed description of the curves is as follows:\\ 
{\it solid black curve}: The calculated $J/\psi$ suppression; \\
{\it solid red curve}: 10\% decrease in $N_{J/\psi(0)}$ and 10\% increase in $N_c(0)$;\\ 
{\it dashed red curve}: 10\% increase in $N_{J/\psi(0)}$ and 10\% decrease in $N_c(0)$;\\ 
{\it solid blue curve}: 10\% decrease in $\Gamma_{diss}$ and 10\% increase in $\Gamma_{recomb}$;\\ 
{\it dashed blue curve}: 10\% increase in $\Gamma_{diss}$ and 10\% decrease in $\Gamma_{recomb}$;\\} 
\label{fig:error_all}
\end{figure}

%gans rev1 error criteria
Many of the above mentioned discrepancies and uncertainties result in an error in either the initial conditions or in the dissociation or recombination rates. In order to estimate the consequence of estimation errors in the initial conditions and the dissociation or recombination rates, the $J/\psi$ suppression is calculated by modifying the initial conditions and $\Gamma_{diss}$ and $\Gamma_{recomb}$ by 10\%. In order to maximize the effect of error in initial conditions, the values of $N_{J/\psi}(0)$ and $N_c(0)$ are modified in opposite directions, i.e. one is decreased by 10\% and the other is increased by 10\% and vice-versa. Similarly, on the dissociation and recombination rates, one is decreased by 10\%, while the other is increased by 10\%.
Figure ~\ref{fig:error_all} depicts the changes in final $J/\psi$ suppression due to various modifications in the initial conditions, $\Gamma_{diss}$ and $\Gamma_{recomb}$. 
%%%   The description of the various curves is as follows:
%%%   \begin{itemize}
%%%   \item solid black curve: The calculated $J/\psi$ suppression.
%%%   \item solid red curve: 10\% decrease in $N_{J/\psi(0)}$ and 10\% increase in $N_c(0)$; 
%%%   \item dashed red curve: 10\% increase in $N_{J/\psi(0)}$ and 10\% decrease in $N_c(0)$; 
%%%   \item solid blue curve: 10\% decrease in $\Gamma_{diss}$ and 10\% increase in $\Gamma_{recomb}$; 
%%%   \item dashed blue curve: 10\% increase in $\Gamma_{diss}$ and 10\% decrease in $\Gamma_{recomb}$; 
%%%   \end{itemize} 
It is interesting to note that the final $J/\psi$ suppression is to some extent resilient to error in initial conditions and dissociation or recombination rates. This can be understood by observing that the coupled rate equations lead to a negative feedback system. For example, if $N_{J/\psi}(0)$ were to increase by 10\%, then a slightly larger number of $J/\psi$ would dissociate into $c\bar{c}$ during the dynamical evolution and thus reducing the $J/\psi$ population. 
%gans rev1 error criteria

\section{Conclusions}
   We have calculated the recombination cross section at various temperatures for recombination of $c\bar{c}$ into $J/\psi$ at both ALICE and CMS experimental conditions using pQCD. A temperature dependent Debye color screened phenomenological potential has been utilized for computing the $J/\psi$ wavefunction which is used in the calculation of the recombination cross-section. 
We have established a set of rate equations which combines gluonic dissociation, collisional damping along with the recombination. CNM effects, namely shadowing is also incorporated in the current work. 
We find that the our final results explain the experimental $J/\psi$ suppression data at ALICE at forward rapidity and CMS at mid rapidity region to a good extent. This indicates that pQCD can be a reasonable approximation for recombination calculations at ALICE forward rapidity and CMS mid rapidity. At ALICE mid rapidity, where the momentum ranges are much lower, the  discrepancy due to pQCD calculations is larger. However, pQCD is able to capture the trend that recombination would be higher at ALICE mid rapidity than at forward rapidity. A more exact estimate of recombination at ALICE mid rapidity would require non-perturbative QCD.  
We have also shown that for charmonium, modification of the formation time due to temperature leads to negligible suppression due to color screening.
 The other causes of the difference between the experimental data and the predicted suppression could possibly be attributed due to the limitations in the accuracy of determining the initial conditions for the rate equations, incomplete thermalization of strange quarks, limitation in the precise estimate of shadowing effect and $J/\psi$ yield during hadronization. 

\section*{Acknowledgment} 
One of the authors (S.G.) acknowledges Broadcom India Research Pvt. Ltd. for allowing the use of its computational resources required for this work. M.M. is grateful to the Department of Science and Technology (DST), New Delhi for financial assistance under the Fast-Track Young Scientist project.

\newpage
\appendix
\section{Feynman Diagrams, Group Theoretic Factors and $\Gamma_{recomb}$}
The Feynman diagrams pertaining to the calculation of $\sigma_{recomb}$ is shown in Fig.~\ref{fig:feyndiag}.

\begin{figure}[h!]
%\begin{figure}
\begin{minipage}[h!]{0.4\textwidth}
\includegraphics[width = 60mm,height = 42mm]{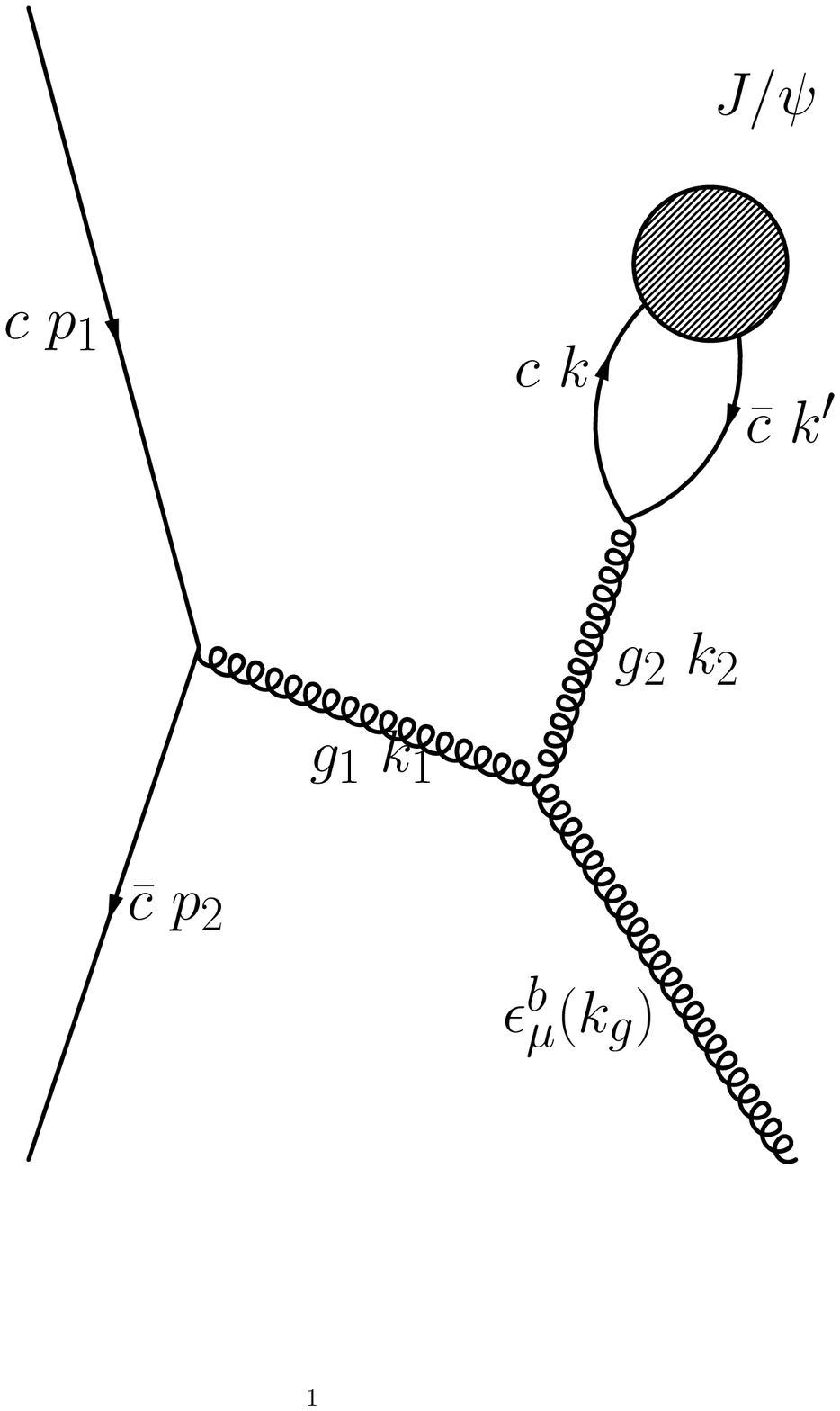} 
%\label{fig:Jpsi1}
\end{minipage}
\hfill
\begin{minipage}[]{0.4\textwidth}
\includegraphics[width = 60mm,height = 42mm]{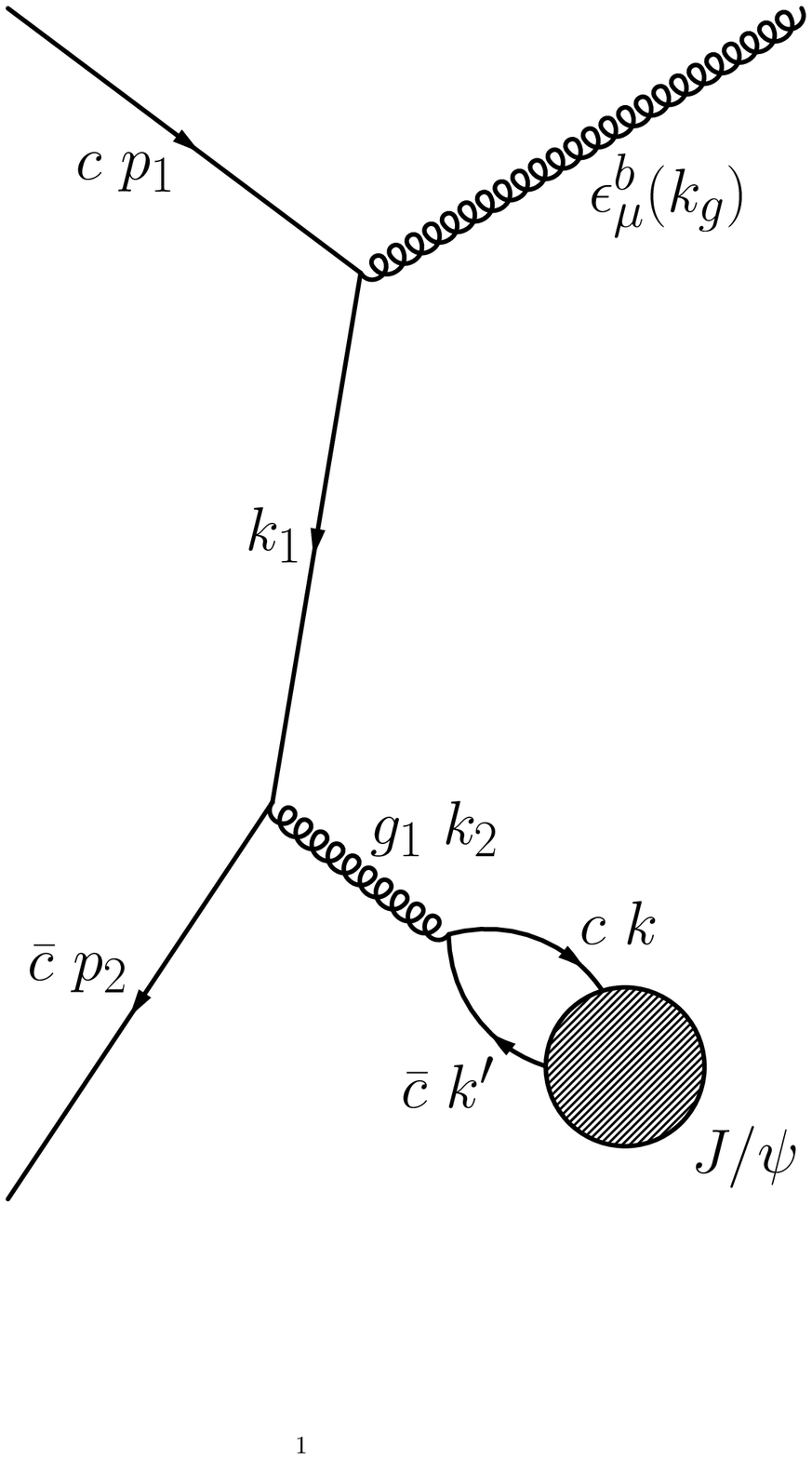}
\end{minipage}

%\hfill
\begin{minipage}[]{0.4\textwidth}
\includegraphics[width = 60mm,height = 42mm]{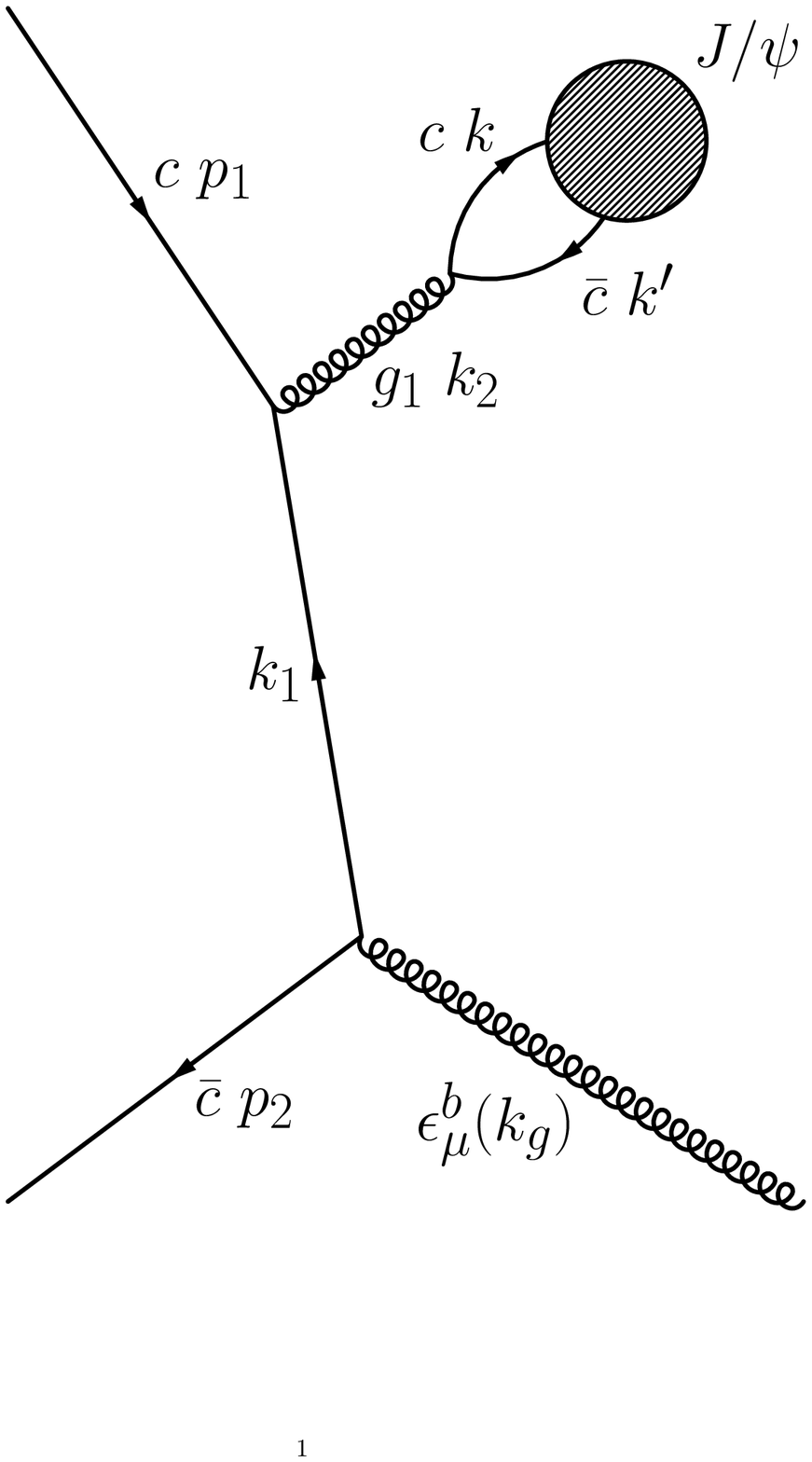}
\label{fig:Jpsi1}
%\caption{roman numerals}
\end{minipage}

%\ContinuedFloat
\begin{minipage}{0.4\textwidth}
\includegraphics[width = 60mm,height = 42mm]{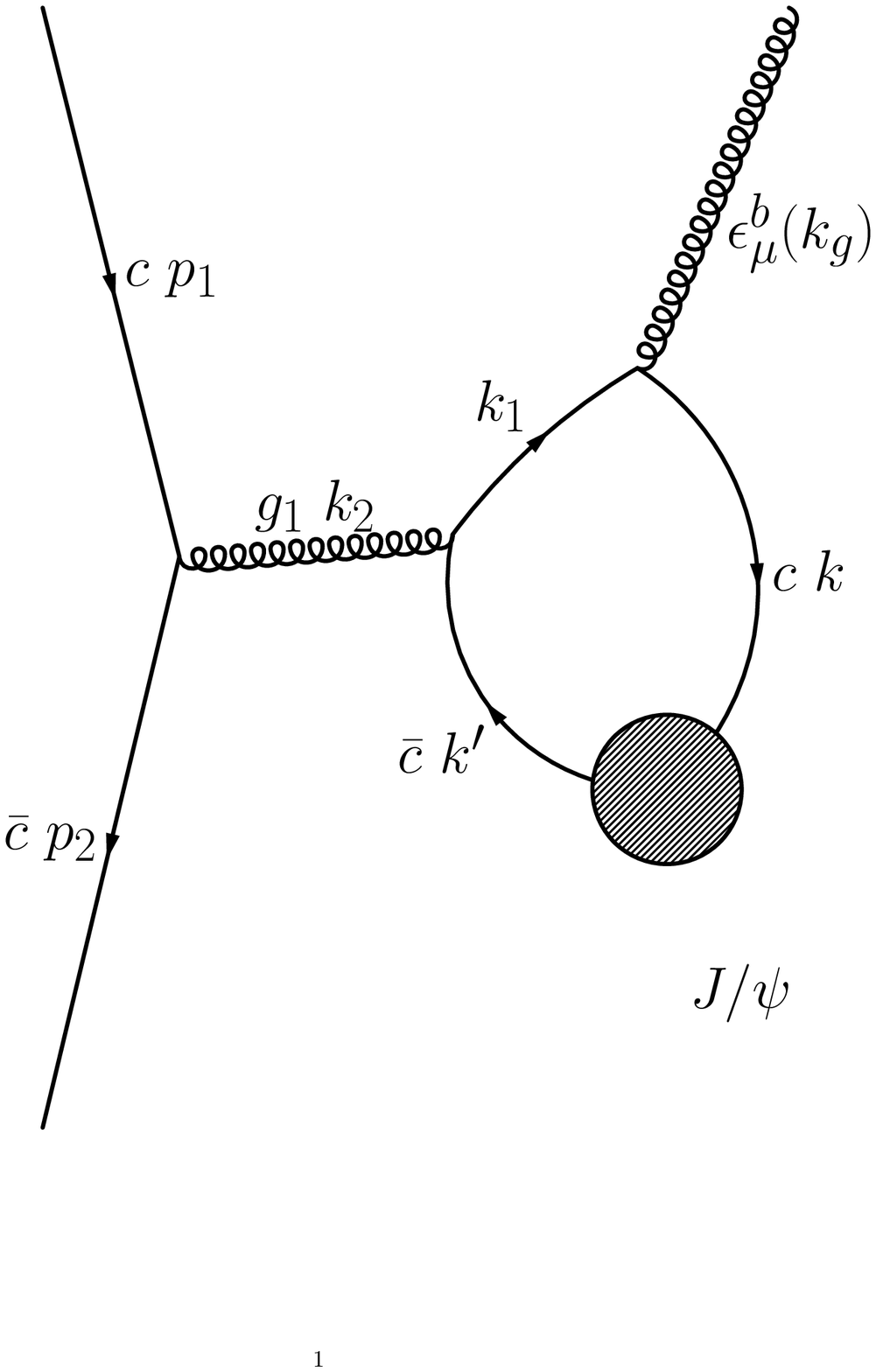}
%\label{fig:Jpsi4}
\end{minipage}

%\phantomcaption
%\caption{iv}
%\begin{subfigure}[h!]
\begin{minipage}{0.4\textwidth}
\includegraphics[width = 60mm,height = 42mm]{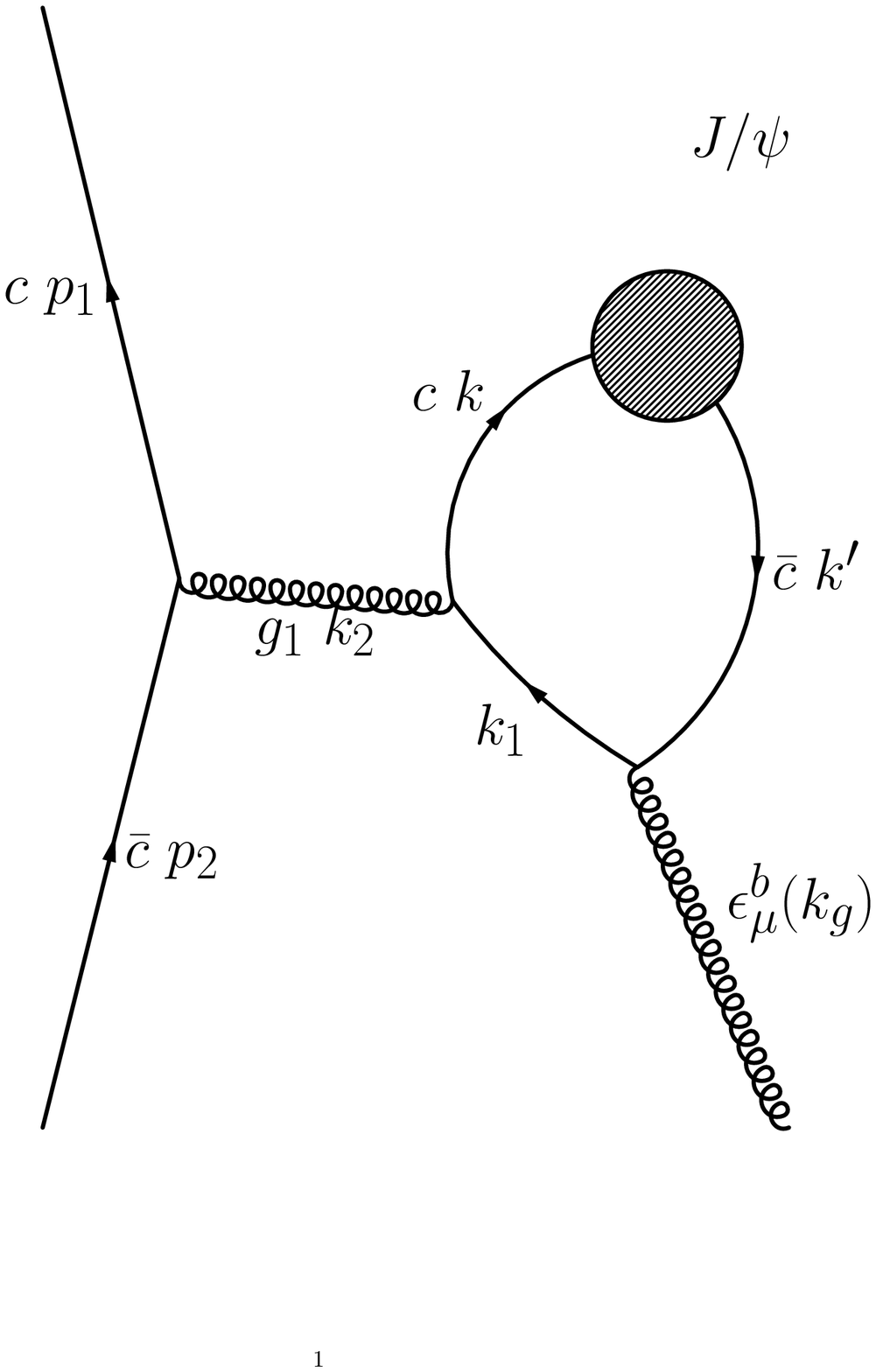}
\end{minipage}
%\end{subfigure}
\end{figure}

\newpage
\begin{figure}[ht]

%\begin{figure}[h!]
\begin{minipage}{0.4\textwidth}
\includegraphics[width = 60mm,height = 41mm]{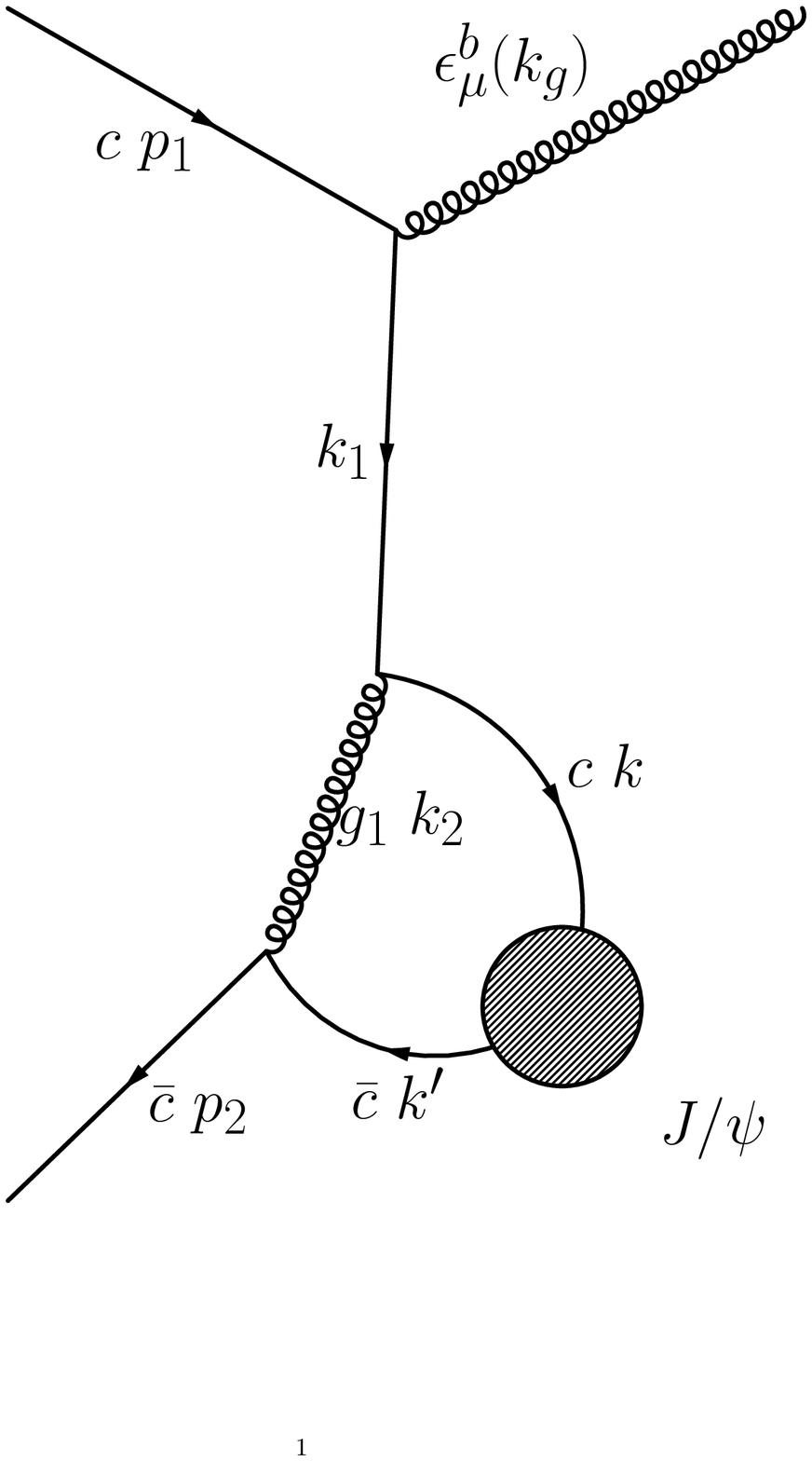}
\end{minipage}

\begin{minipage}{0.4\textwidth}
\includegraphics[width = 60mm,height = 41mm]{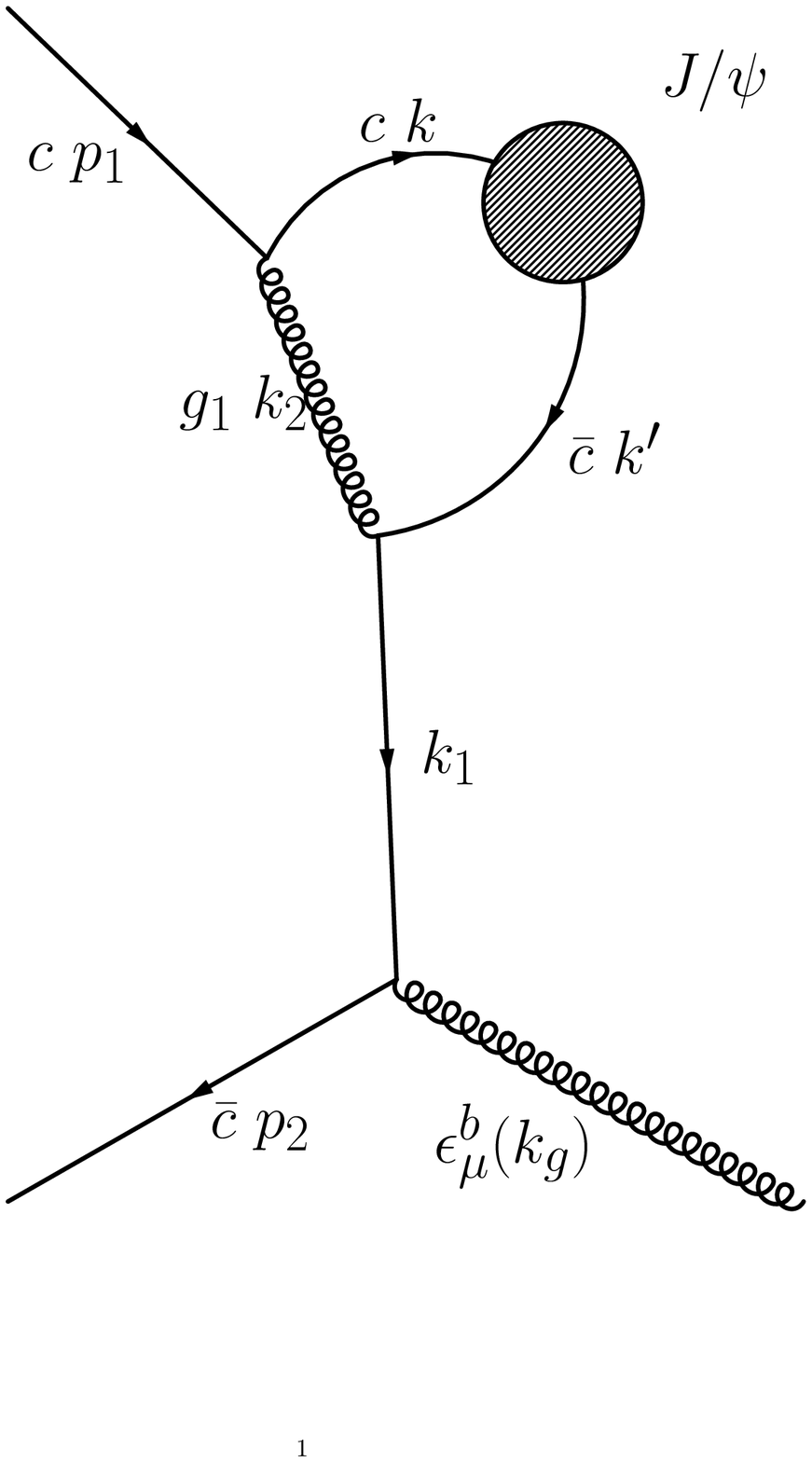}
%\label{fig:Jpsi7}
\end{minipage}
\hfill
\begin{minipage}{0.4\textwidth}
\includegraphics[width = 60mm,height = 41mm]{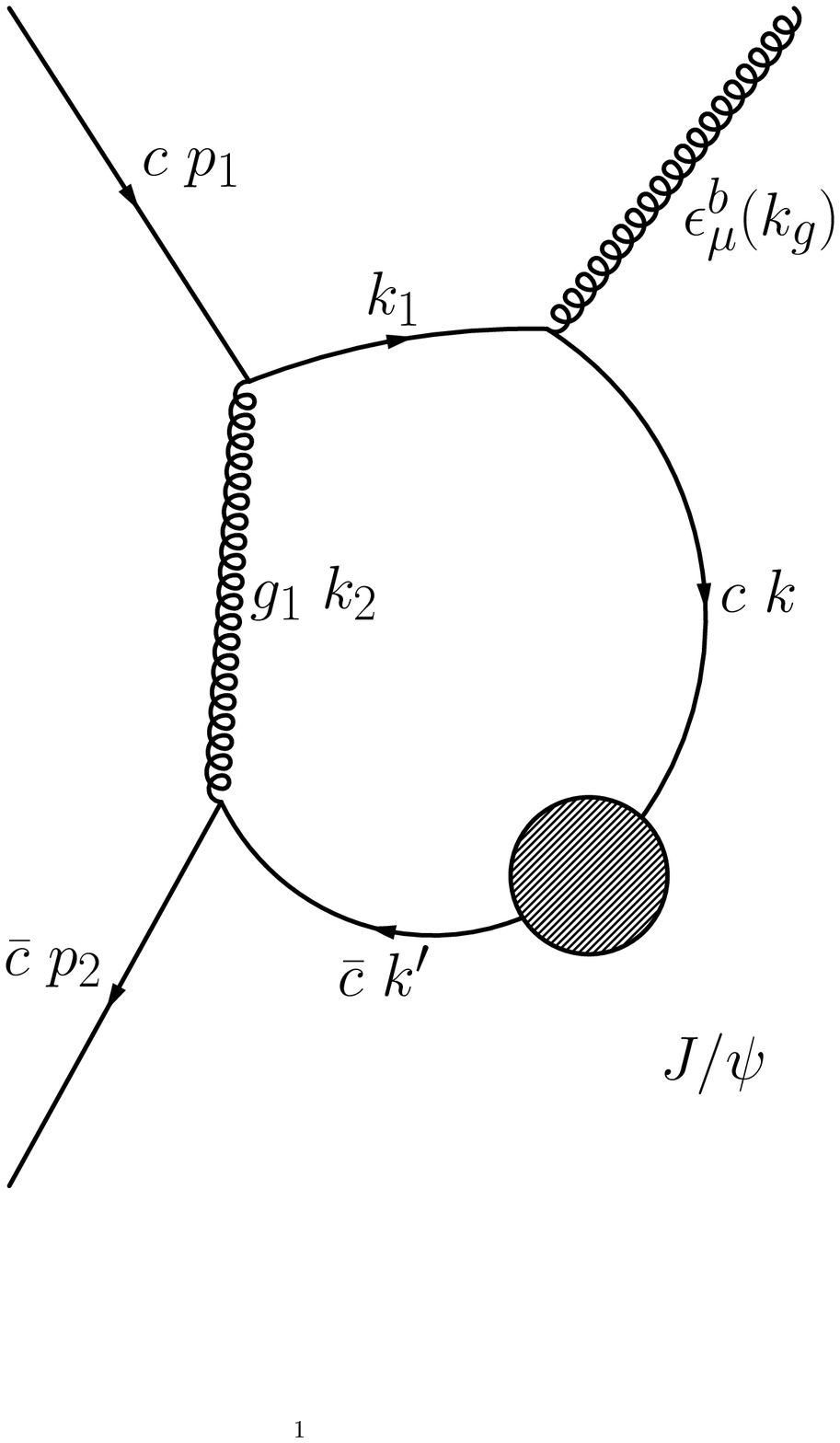}
\end{minipage}
%\end{figure}
%\newpage
\begin{minipage}{0.4\textwidth}
\includegraphics[width = 60mm,height = 41mm]{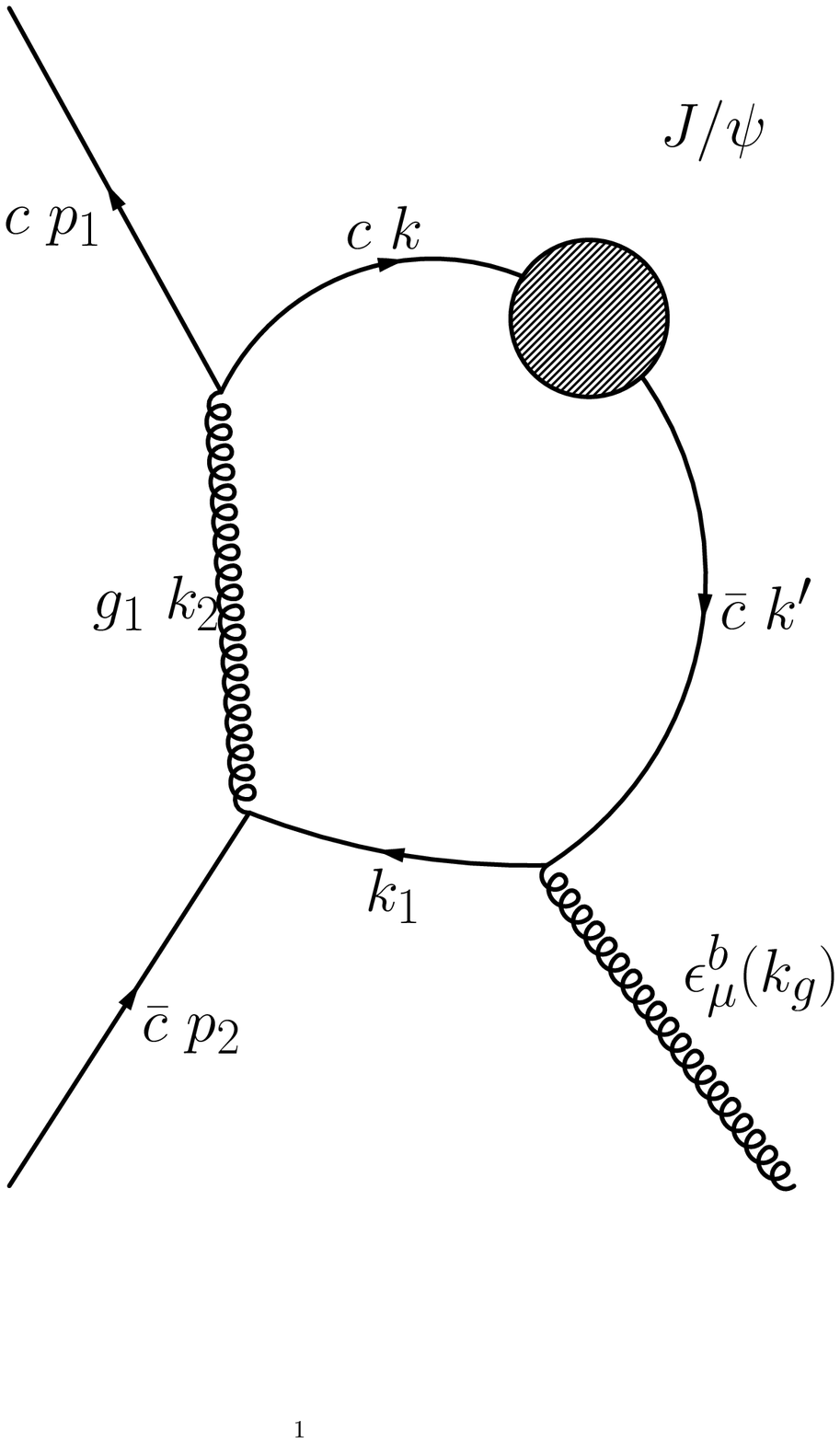}
%\label{fig:Jpsi9}
\end{minipage}
%\caption{ix}
\hfill
\begin{minipage}{0.4\textwidth}
\includegraphics[width = 60mm,height = 41mm]{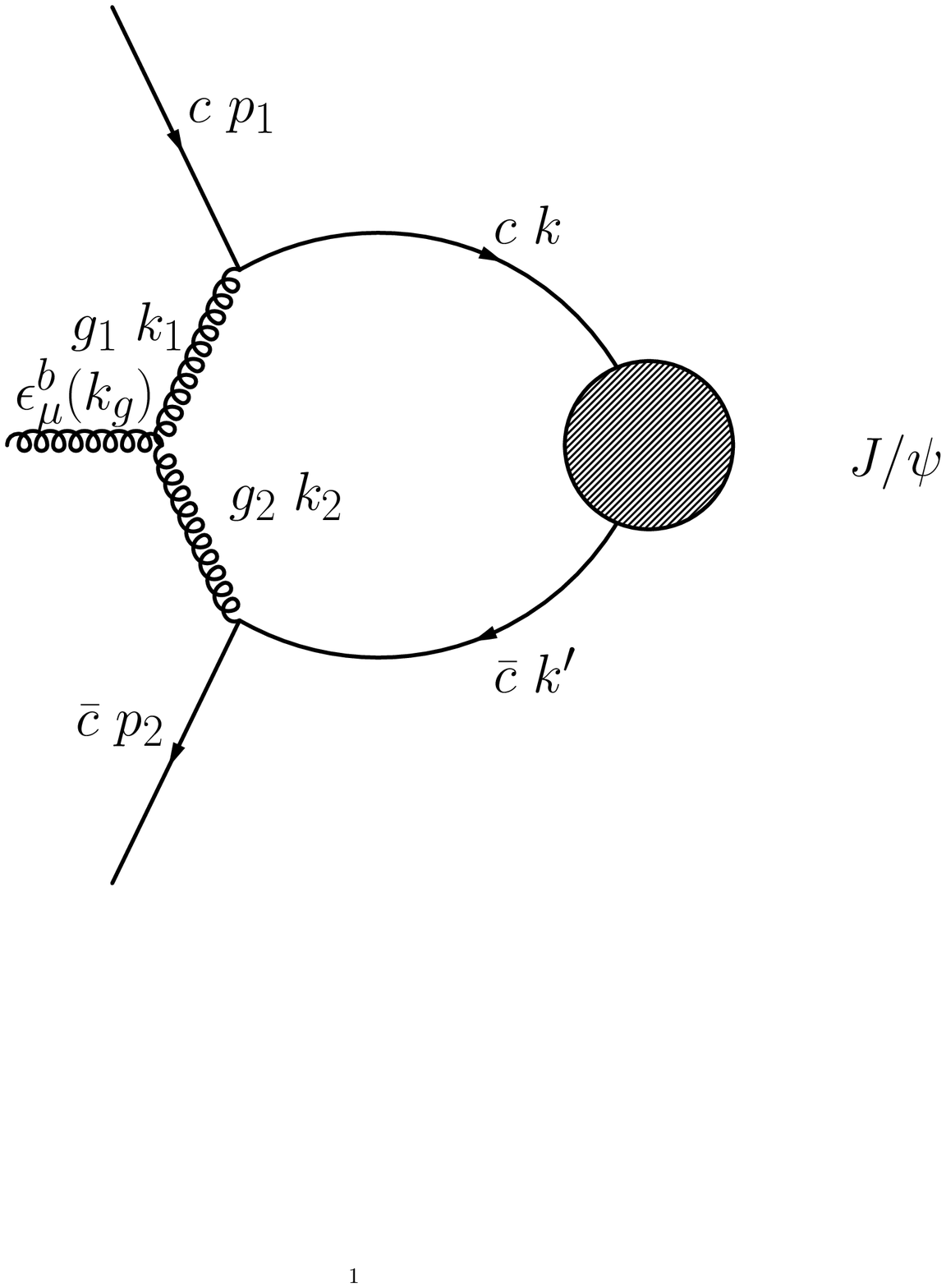}
%\label{fig:Jpsi10}
\end{minipage}
%\caption{$c\bar{c}$ scattering to emit a gluon and $J/\psi$.}
\caption{$c\bar{c}$ scattering to emit a gluon and $J/\psi$.}
\label{fig:feyndiag}
\end{figure}

%%%%%%%%%%%%%%%%%%%%%%%%%%%%%%%%%%%%%%%%%%%%%%%%%%%5

\newpage
The group theoretic color factors appearing in the calculation of the above diagrams is given below. In the equations written below, $G_{i,j}$ refers to the group theoretic color factor appearing in the calculation of $M^*_i\,M_j$. $N_c$ = number of colors = 3. the value of $h$ is given by
\begin{math}
h^{abc} = d^{abc} + if^{abc}
\end{math}
where f is the totally antisymmetric structure constants of SU(3), while d is totally symmetric. 
\vskip 0.15in
\begin{math}
%G_{1,1}=1/4*(N_c)^3-1/4*(N_c); %G_{1,2}=1/8*i*N_c^3-1/8*i*N_c;
G_{1,1}=N_c/4(N_c^2-1); G_{1,2} = i\,N_c/8(N_c^2-1);
\end{math}
\\
\begin{math}
%G_{1,3}=-1/8*i*N_c^3+1/8*i*N_c; %G_{1,4}=1/8*i*N_c^3-1/8*i*N_c;
G_{1,3} = iN_c/8(1-N_c^2); G_{1,4} = i\,N_c/8(N_c^2-1);
\end{math}
\\
\begin{math}
%G_{1,5} = 1/8*i*N_c^3+1/8*i*N_c; %G_{1,6} = 1/8*i*N_c^2+1/8*i;
G_{1,5} = iN_c/8(N_c^2+1); G_{1,6} = i\,/8(N_c^2+1);
\end{math}
\\
\begin{math}
%G_{1,7} = 1/8*i*N_c^2-1/8*i; %G_{1,8} = 1/8*i*N_c^2+1/8*i;
G_{1,7} = i/8(N_c^2-1); G_{1,8} = i\,/8(N_c^2+1);
\end{math}
\\
\begin{math}
%G_{1,9} = 1/8*i*N_c^2-1/8*i; %G_{1,10} = 2*i*fffh-1/8*N_c^4+1/8*N_c^2;
G_{1,9} = i/8(N_c^2-1); 
\end{math}
\begin{math}
G_{1,10} = 2\,i.f^{ba'c'}f^{bac}f^{a'ae}h^{c'ce}+N_c^2/8(1-N_c^2);
\end{math}
\\
\begin{math}
%G_{2,2} = 1/8*N_c^3-1/4*N_c+1/8/N_c; %G_{2,3} = 1/8*N_c+1/8/N_c;
G_{2,2} = N_c^3/8-N_c/4+1/8N_c; G_{2,3} = 1/8(N_c+1/N_c);
\end{math}
\\
\begin{math}
G_{2,4} = 4h^{a'ba}h^{a'ab}; 
G_{2,5} = 4h^{ba'a}h^{ba'a};
\end{math}
\\
\begin{math}
%G_{2,6} = 1/8*N_c^2+1/4-1/8/N_c^2; %G_{2,7} = 1/8-1/8/N_c^2;
G_{2,6} = N_c^2/8+1/4-1/8N_c^2; G_{2,7} = 1/8(1-1/N_c^2);
\end{math}
\\
\begin{math}
%G_{2,8} = 1/8-1/8/N_c^2; %G_{2,9} = 1/8*N_c^2-1/8/N_c^2;
G_{2,8} = 1/8(1-1/N_c^2); G_{2,9} = 1/8(N_c^2-1/N_c^2);
\end{math}
\\
\begin{math}
%G_{2,10} = 1/8*i*N_c^2-1/8*i;
G_{2,10} = i/8(N_c^2-1);
\end{math}
\\
\begin{math}
%G_{3,3} = 1/8*N_c^3-1/4*N_c+1/8/N_c;
G_{3,3} = N_c^3/8-N_c/4+1/8N_c; 
\end{math}
\begin{math}
G_{3,4} = 4h^{aa'b}h^{aa'b};
\end{math}
\\
\begin{math}
G_{3,5} = 4h^{ba'a}h^{baa'}n;
\end{math}
\\
\begin{math}
%G_{3,6} = 1/8-1/8/N_c^2; %G_{3,7} = 1/8*N_c^2+1/4-1/8/N_c^2;
G_{3,6} = 1/8(1-1/N_c^2); G_{3,7} = N_c^2/8+1/4-1/8N_c^2;
\end{math}
\\
\begin{math}
%G_{3,8} = 1/8*N_c^2-1/8/N_c^2; %G_{3,9} = 1/8-1/8/N_c^2;
G_{3,8} = 1/8(N_c^2-1/N_c^2); G_{3,9} = 1/8(1-1/N_c^2);
\end{math}
\\
\begin{math}
%G_{3,10} = 1/8*i*N_c^2-1/8*i;
G_{3,10} = i/8(N_c^2-1);
\end{math}
\\
\begin{math}
%G_{4,4} = 1/8*N_c^3-1/4*N_c+1/8/N_c; %G_{4,5} = 1/8*N_c+1/8/N_c;
G_{4,4} = N_c^3/8-N_c/4+1/8N_c; G_{4,5} = 1/8(N_c+1/N_c);
\end{math}
\\
\begin{math}
%G_{4,6} = 1/8-1/8/N_c^2; %G_{4,7} = 1/8*N_c^2+1/4-1/8/N_c^2;
G_{4,6} = 1/8(1-1/N_c^2); G_{4,7} = N_c^2/8+1/4-1/8N_c^2;
\end{math}
\\
\begin{math}
%G_{4,8} = 1/8*N_c^2+1/4-1/8/N_c^2; %G_{4,9} = 1/8-1/8/N_c^2;
G_{4,8} = N_c^2/8+1/4-1/8N_c^2; G_{4,9} = 1/8(1-1/N_c^2);
\end{math}
\\
\begin{math}
%G_{4,10} = 1/8*i*N_c^2+1/8*i;
G_{4,10} = i/8(N_c^2+1);
\end{math}
\\
\begin{math}
%G_{5,5} = 1/8*N_c^3-1/4*N_c+1/8/N_c; %G_{5,6} = 1/8*N_c^2-1/8/N_c^2;
G_{5,5} = N_c^3/8-N_c/4+1/8N_c; G_{5,6} = 1/8(N_c^2-1/N_c^2);
\end{math}
\\
\begin{math}
%G_{5,7} = 1/8-1/8/N_c^2; %G_{5,8} = 1/8-1/8/N_c^2;
G_{5,7} = 1/8(1-1/N_c^2); G_{5,8} = 1/8(1-1/N_c^2);
\end{math}
\\
\begin{math}
%G_{5,9} = 1/8*N_c^2+1/4-1/8/N_c^2; %G_{5,10} = 1/8*i*N_c^2+1/8*i;
G_{5,9} = N_c^2/8+1/4-1/8N_c^2; G_{5,10} = i\,/8(N_c^2+1);
\end{math}
\\
\begin{math}
%G_{6,6} = 1/8*N_c^3-1/4*N_c+1/8/N_c;
G_{6,6} = N_c^3/8-N_c/4+1/8N_c; 
\end{math}
\begin{math}
G_{6,7} = 4h^{a'ab}h^{a'ba};
\end{math}
\\
\begin{math}
%G_{6,8} = 1/8*N_c+1/8/N_c;
G_{6,8} = 1/8(N_c+1/N_c); 
\end{math}
\begin{math}
G_{6,9} = 4h^{ba'a}h^{ba'a};
\end{math}
\\
\begin{math}
%G_{6,10} = 1/8*i*N_c^3-1/8*i*N_c;
G_{6,10} = iN_c/8(N_c^2-1);
\end{math}
\\
\begin{math}
%G_{7,7} = 1/8*N_c^3-1/4*N_c+1/8/N_c;
G_{7,7} = N_c^3/8-N_c/4 + 1/8N_c; 
\end{math}
\begin{math}
G_{7,8} = 4h^{aa'b}h^{aa'b};
\end{math}
\\
\begin{math}
%G_{7,9} = 1/8*N_c+1/8/N_c; %G_{7,10} = 1/8*i*N_c^3+1/8*i*N_c;
G_{7,9} = 1/8(N_c+1/N_c); G_{7,10} = i\,N_c/8(N_c^2+1);
\end{math}
\\
\begin{math}
%G_{8,8} = 1/8*N_c^3-1/4*N_c+1/8/N_c;
G_{8,8} = N_c^3/8-N_c/4+1/8N_c; 
\end{math}
\begin{math}
G_{8,9} = 4h^{a'ba}h^{ba'a};
\end{math}
\\
\begin{math}
%G_{8,10} = 1/8*i*N_c^3-1/8*i*N_c;
G_{8,10} = i\,N_c/8(N_c^2-1);
\end{math}
\\
\begin{math}
%G_{9,9} = 1/8*N_c^3-1/4*N_c+1/8/N_c; %G_{9,10} = 1/8*i*N_c^3-1/8*i*N_c;
G_{9,9} = N_c^3/8-N_c/4+1/8N_c; G_{9,10} = i\,N_c/8(N_c^2-1);
\end{math}
\\
\begin{math}
%G_{10,10} = 1/4*(N_c)^3-1/4*(N_c);
G_{10,10} = N_c/4(N_c^2-1);
\end{math}
\\
\section{Infrared divergence of Fermion propagator}
	This section proves the cancellation of the infrared divergence due to the fermion propagator. 
When the emitted gluon momentum approaches 0, the $J/\psi$ center of mass reference frame becomes equivalent to the $c\bar{c}$ center of mass reference frame. The two momenta $\vec{p}_1$ and $\vec{p}_2$ becomes equal and opposite. With this condition, let's consider the sum of the Feynman diagrams 6 and 7.
\\
\begin{eqnarray}
\nonumber M_6 = \displaystyle\int \Bigg ( \left ( \bar{v}(p_2)(ig\gamma_{\nu}t^a)v(k') (\frac{-ig^{\nu\sigma}}{k_2^2}) \right )\\  
\nonumber \Big  ( \bar{u}(k)(ig\gamma_{\sigma}{t^a})\frac{i}{\slashed{k_1}-m}(ig\gamma_{\mu}t^b)u(p_1)\Big )\psi_T(k,k') \Bigg ) \\
dk\,\epsilon^{\mu,b}(k_g)~
\end{eqnarray}

\begin{eqnarray}
\nonumber M_7 = \displaystyle\int \left ( \bar{v}(p_2)(ig\gamma_{\mu}t^b)\frac{i}{\slashed{k_1}-m}(ig\gamma_{\nu}t^a)v(k') \right )\\  
(\frac{-ig^{\nu\sigma}}{k_2^2}) 
\nonumber \Big ( \left ( \bar{u}(k)(ig\gamma_{\sigma}{t^a})u(p_1)\psi_T(k,k')\right )\Big ) \\
dk\,\epsilon^{\mu,b}(k_g)~
\end{eqnarray}
%which becomes
%\begin{eqnarray}
%\nonumber M_6 = \displaystyle\int \Bigg ( \left ( \bar{v}(p_2)(ig\gamma_{\nu}t^a)v(k') (\frac{-i}{k_2^2}) \right ) \\
% \Big ( \bar{u}(k)(ig\gamma^{\nu}{t^a})\frac{i}{\slashed{k_1}-m}(ig\gamma_{\mu}t^b)u(p_1)\Big )\psi_T(k,k') \Bigg ) \\
%dk\,\epsilon^{\mu,b}(k_g)
%\end{eqnarray}
%
%\begin{eqnarray}
%\nonumber M_7 = \displaystyle\int \Big ( \bar{v}(p_2)(ig\gamma_{\mu}t^b)\frac{i}{\slashed{k_1}-m}(ig\gamma_{\nu}t^a)v(k') \Big ) (\frac{-i}{k_2^2}) \\
%\Big ( \left ( \bar{u}(k)(ig\gamma^{\nu}{t^a})u(p_1)\psi_T(k,k')\right ) \Big ) \\
%dk\,\epsilon^{\mu,b}(k_g)
%\end{eqnarray}
\\
\\
After contracting the $\sigma$ index, the $M_6 +M_7$ integrand would contain the expression\\
\begin{eqnarray}
\nonumber  \Bigg ( \Big ( \bar{v}(p_2)(ig\gamma_{\nu}t^a)v(k')  \Big ) \\
 \left ( \bar{u}(k)(ig\gamma^{\nu}{t^a})\frac{i}{\slashed{k_1}-m}(ig\gamma_{\mu}t^b)u(p_1)\right ) \Bigg ) \\
+ \\ 
\nonumber  \Bigg ( \left ( \bar{v}(p_2)(ig\gamma_{\mu}t^b)\frac{i}{\slashed{k_1}-m}(ig\gamma_{\nu}t^a)v(k') \right ) \\ 
\Big ( \bar{u}(k)(ig\gamma^{\nu}{t^a})u(p_1) \Big )\Bigg )
\end{eqnarray}

The first term is from the expression for $M_6$, while the second is from the $M_7$ expression. The gluon polarization term is reintroduced in the end.
The term $\frac{i}{\slashed{k}_1-m}$ = $\frac{i(\slashed{k}_1+m)}{k_1^2 - m^2}$ =$\frac{i\slashed{k}_g}{2p_x.k_g} - \frac{i(\slashed{p}_x+m)}{2p_x.kg}$.
The first term will have no infrared divergence as $k_g \rightarrow 0$, and only the second term needs to be taken care of. Here $p_x$ is either $p_1$ for the expression for $M_6$, or $p_2$ in the case of $M_7$. 

With the first term removed, the non scalar terms in the expression consist of
%modify the below expression
\begin{eqnarray}
\nonumber  \Bigg ( \Big ( \bar{v}(p_2)(ig\gamma_{\nu}t^a)v(k')  \Big ) \\
 \left ( \bar{u}(k)(ig\gamma^{\nu}{t^a})\frac{\slashed{p}_1 +m}{2{p_1.k_g}}(ig\gamma_{\mu}t^b)u(p_1)\right ) \Bigg )  + \\
\nonumber  \Bigg ( \left ( \bar{v}(p_2)(ig\gamma_{\mu}t^b)\frac{\slashed{p}_2 + m}{2p_2.k_g}(ig\gamma_{\nu}t^a)v(k') \right ) \\
\Big ( \bar{u}(k)(ig\gamma^{\nu}{t^a})u(p_1) \Big )\Bigg )
\end{eqnarray}
%The overall negative sign from $-(\slashed{p}_x + m)$ is not shown as its common to both terms.
From straightforward Dirac algebra, this is simplified to\\ 
\begin{eqnarray}
\nonumber  \Bigg ( \Big ( \bar{v}(p_2)(ig\gamma_{\nu}t^a)v(k')  \Big ) \\
 \left ( \bar{u}(k)(ig\gamma^{\nu}{t^a})\frac{2p_{1\mu}}{2{p_1}.k_g}(igt^b)u(p_1)\right ) \Bigg )  + \\
\nonumber  \Bigg ( \bar{v}(p_2)(igt^b)\frac{2p_{2\mu}}{2p_2.k_g}(ig\gamma_{\nu}t^a)v(k') \\
 \left ( \bar{u}(k)(ig\gamma^{\nu}{t^a})u(p_1) \right )\Bigg )
\end{eqnarray}

To see the infrared behavior, we write the particle spinor in terms of the antiparticle spinor and vice versa in expression for $M_6$ i.e. the first expression.\\ 
\begin{eqnarray}
\nonumber v(k') = -i\gamma_2u^*(k')\\
\nonumber \bar{v}(p_2) = -i\bar{u}^*(p_2)\gamma_2
\end{eqnarray}
\\
We get
\begin{eqnarray}
\bar{v}(p_2)\gamma_{\nu}v(k') \\
\nonumber= -i\bar{u}^*(p_2)\gamma_2 \gamma_{\nu} (-i) \gamma_2 u^*(k')\\
\nonumber= -(-1)^{\nu \ne 2} \bar{u}^*(p_2) \gamma_{\nu} u^*(k')\\
\nonumber= -(-1)^{\nu \ne 2} \bar{u}^*(p_2) (-1)^{\nu=2}\gamma^*_{\nu} u^*(k')\\
\nonumber=  \left ( \bar{u}(p_2)\gamma_{\nu} u(k')\right )^*\\
%= - \left ( \bar{v}(p_2)\gamma_{\nu} u(k')\right )
\end{eqnarray}
%The last equality can be seen from the fact that with weyl representation, $\left ( \bar{v(p_2)}\gamma_{\nu} u(k')\right )$ is purely real.
The various minus signs used in arriving at B11 are as follows
\begin{enumerate}[(a)]
\item $-1^{\nu \ne 2}$: from the commutation relation of $\gamma_{\nu}$ and $\gamma_2$ 
\item $-$ : from $\gamma_2^2 = -1$
\item $-1^{\nu = 2}$: from $\gamma_{2}$ = $-\gamma_2^*$ 
\end{enumerate}

On similar lines
\begin{equation}
\nonumber \left ( \bar{u}(k)(\gamma^{\nu})\frac{2p_{1\mu}}{2p_1.k_g}u(p_1)\right ) 
\end{equation}

\begin{equation}
=\frac{2p_{1\mu}}{2p_1.k_g}\left ( \bar{v}(k)(\gamma^{\nu})v(p_1)\right )^*
\end{equation}
\\
\\
Taking the product of the two expressions B11 and B12, we get:

\begin{equation}
=  \left ( \bar{u}(p_2)\gamma_{\nu} u(k')\right )^*
\frac{2p_{1\mu}}{2p_1.k_g}\left ( \bar{v}(k)(\gamma^{\nu})v(p_1)\right )^*
\end{equation}

This gives the expression for incoming particles of momenta $k$ and $k'$ and outgoing particles of velocity $p_1$ and $p_2$.
The  conjugate transpose  relates the above expression to the amplitude M for incoming particles of momenta $p_1$ and $p_2$ and outgoing particles of momenta $k$ and $k'$ .
Hence the above expression becomes, 
\begin{eqnarray}
\nonumber=  \left ( \bar{u}(k')\gamma_{\nu} u(p_2)\right )^t
\frac{2p_{1\mu}}{2p_1.k_g}\left ( \bar{v}(p_1)(\gamma^{\nu})v(k)\right )^t\\
\nonumber=  \left ( \bar{u}(k')\gamma_{\nu} u(p_2)\right )
\frac{2p_{1\mu}}{2p_1.k_g}\left ( \bar{v}(p_1)(\gamma^{\nu})v(k)\right )\\
= \frac{2p_{1\mu}}{2p_1.k_g}\left [ \left ( \bar{u}(k')\gamma_{\nu} u(p_2)\right )
\left ( \bar{v}(p_1)(\gamma^{\nu})v(k)\right ) \right ]
\end{eqnarray}

The term in the square brackets is an inner product which would be an invariant to rotation in 3D space. Rotation of $\pi$ radians along a suitable axis interchanges $p_1$ and $p_2$ as well as $k$ and $k'$.

\begin{equation}
= \frac{2p_{1\mu}}{2p_1.k_g} \left [ \left ( \bar{u}(k)\gamma_{\nu} u(p_1)\right )
\left ( \bar{v}(p_2)(\gamma^{\nu})v(k')\right ) \right ]
\end{equation}

This expression for $M_6$ is now identical to the expression for  $M_7$ except for the scalar terms $\frac{p_{1\mu}}{2p_1.k_g}$ and $\frac{p_{2\mu}}{2p_2.k_g}$ 
Adding $M_6$ and $M_7$ and reintroducing the gluon polarization term, we get,
\\
\begin{equation}
(...) (\frac{p_{1\mu}\epsilon^{\mu}(k_g)}{p_1.k_g} + \frac{p_{2\mu}\epsilon^{\mu}(k_g)}{p_2.k_g})
\end{equation}
\\
Since we are considering only transverse polarization of the gluon, the time component is 0 and noting that $\vec{p}_1$ = -$\vec{p}_2$, we get,
\begin{equation}
\label{eq:infraredcancel}
(...) (\frac{p_{1i}\epsilon^{i}(k_g)}{p_1.k_g} - \frac{p_{1i}\epsilon^{i}(k_g)}{p_2.k_g})
\end{equation}
with i = 1,2,3.
\\
In the center of mass frame, if the gluon 3 momentum $\vec{k}_g$ is perpendicular to the 3 momentum $\vec{p}_1$ and $\vec{p}_2$, then $p_1.k_g$ = $p_2.k_g$ = $E_c\,E_g$, where $E_c$ = charm quark energy and $E_g$ = gluon energy in the center of mass frame. With this, the above expression evaluates to 0. 
	When the gluon 3 momentum is parallel to $p_1$, $\epsilon(k_g)$ would be perpendicular to the 3 momentum $\vec{p}_1$, and the numerator vanishes.
	Thus, if the gluon 3 momentum is perpendicular or parallel to charm quark (anti-quark) 3 momentum, the infrared divergence cancels.
In other inclinations, there would be partial cancellation of varying degrees. 
%If we pair terms in which the $M_6$ and $M_7$ diagrams have opposite gluon momentum, $p_1.(k_{g_0}, \vec{k}_g)$ = $p_2.(k_{g_0}, -\vec{k}_g)$ and the above term evaluates to 0.
%This proves that infrared divergence for the fermion propagator cancels between $M_6$ and $M_7$. Similar proof can be obtained for infrared cancellation due to the fermion propagator between $M_2$ and $M_3$, $M_4$ and $M_5$ and finally $M_8$ and $M_9$.
%Hence the infrared divergence terms cancel each other.

	A similar analysis can be done for diagrams 2 and 3, 4 and 5 and finally 8 and 9.

%The temperature dependent values of $\Gamma_{recomb}$ at momentum range corresponding to ALICE and CMS experimental data is shown in table I. 
%\begin{table}
%\caption {Temperature dependent values of $\Gamma_{recomb}$} 
%\begin{tabular}{||l|c|c|c|c|c|c|c|c|c||}
%
%\hline
%
%    Temperature (MeV) &0 &  50 &  100 &  150 &  200 &  250 &  300 & 350 &  368\\
%\hline
%
%    $p_T < 8GeV$(mb) & 1.4401 &  1.2700  & 1.0440 &  0.7980 &  0.6480  & 0.4533  & 0.2510 &  0.0883  & 0.0557\\
%\hline
%   $6.5 < p_T < 30GeV$ (mb) & 0.0528  & 0.0470  & 0.0392 &  0.0304 &  0.0252  & 0.0178 &  0.0099 & 0.0035 &  0.0022\\
%\hline
%\end{tabular}
%\end{table}


\begin{thebibliography}{99}
\bibitem{PKSMCA} M. C. Abreu et al., (NA50 Collaboration), Phys. Lett. B {\bf 477}, 28 (2000); B. Alessandro et al., (NA50 Collaboration), Eur. Phys. J. C {\bf 39}, 335 (2005).
\bibitem{PKSRAR} R. Arnaldi et al., (NA60 Collaboration), Phys. Rev. Lett. {\bf 99}, 132302 (2007); R. Arnaldi (NA60 Collaboration), Presentation at the ECT workshop on "Heavy Quarkonia Production in Heavy-Ion Collisions," Trent (Italy), May 25-29 (2009).
\bibitem{PKSAAD} A. Adare et al., (PHENIX Collaboration), Phys. Rev. Lett. {\bf 98}, 232301 (2007).
\bibitem{ALICEfor} ALICE collaboration, Phys. Rev. Lett. 109 (2012) 072301; arxiv:1202.1383v2 [hep-ex] (2012).
\bibitem{ALICEjpsi} Enrico Scomparin (for ALICE collaboration) Nucl. Phys. A 904-905, 202c-209c (2013); 
\bibitem{PKSCMS1} The CMS Collaboration, J. High Energy Phys. {\bf 05}, 063 (2012).
\bibitem{PKSBAB} B. Abelev et al., (ALICE Collaboration), Phys. Rev. Lett. {\bf 109}, 072301 (2012).
\bibitem{PKSSPS1} J. P. Blaizot and J. Y. Ollitrault, Phys. Rev. Lett. {\bf 77}, 1703 (1996).
\bibitem{PKSSPS2} J. P. Blaizot, P. M. Dinh, and J. Y. Ollitrault, Phys. Rev. Lett. {\bf 85}, 4012 (2000). 
\bibitem{PKSSPS3} A. Capella, E. G. Ferreiro, and A. B. Kaidalov, Phys. Rev. Lett. {\bf 85}, 2080 (2000).
\bibitem{PKSSPS4} A. K. Chaudhuri, Phys. Rev. C {\bf 64}, 054903 (2001); Phys. Lett. B {\bf 527}, 80 (2002).
\bibitem{PKSSPS5} A. K. Chaudhuri, Phys. Rev. Lett. {\bf 88}, 232302 (2002).
\bibitem{PKSSPS6} A. K. Chaudhuri, Phys. Rev. C {\bf 66}, 021902 (2002).
\bibitem{PKSSPS7} A. K. Chaudhuri and P. P. Bhaduri, arXiv:1202.3291[nucl-th], (2012).
\bibitem{yunpen} Yunpeng Liu et al., J. Phys. G {\bf 37}, 075110 (2010).
\bibitem{zhen}   Zhen Qu et al., Nucl. Phys. A 830, 335c (2009).
\bibitem{rishi}  Rishi Sharma and Ivan Vitev, Phys. Rev. C {\bf 87}, 044905 (2013).
%\bibitem{ALICEref} ALICE collaboration, Physics Letters B, 734, 314-327, (2014).
\bibitem{stathadref} A. Andronic, P Braun-Munzinger, K. Redlich, J. Stachel, J of Physics G, {\bf 38}, 124081, (2011).
\bibitem{twice1ref} Yunpeng Liu, Zhen Qu, Nu Xu and Pengfei Zhuang,  Physics Leeters B 678 (2009); arxiv:0907.2723v1 [nucl-th] (2009).
\bibitem{trans2ref} Xingbo Zhao, Ralf Rapp, Nuclear Physics A 859 (2011)
%\bibitem{initial}  The CMS Collaboration, arXiv:1201.5069v2 [nucl-ex], (2013)
\bibitem{rituraj} Captain R. Singh, P. K. Srivastava, S. Ganesh, M. Mishra,  Phys. Rev. C {\bf 92}, 034916 (2015).
\bibitem{pp7TeV} ALICE collaboration, Eur. Phys. J. C {\bf 74}, 2974, (2014); arxiv:1403.3648v2 [nucl-ex].
\bibitem{Wolschin} F. Nendzig, G. Wolschin, Phys. Rev. C {\bf 87}, 024911 (2013).
\bibitem{peskin} G. Bhanot and M. E. Peskin, Nuclear Physics B, {\bf 156}, 391, 1979
\bibitem{Madhu2} P. K.Srivastava, M. Mishra and C. P. Singh, Phys. Rev. C {\bf 87}, 034903 (2013).
\bibitem{gans2} S. Ganesh and M. Mishra, Phys. Rev. C {\bf 91}, 034901 (2015).
\bibitem{vogt} R. Vogt, Phys. Rev. C {\bf 81}, 044903 (2010).
%\bibitem{cnmel0} I. Vitev, T. Goldman, M. B. Johnson, J. W. Qiu, Phys. Rev. D {\bf 74}, 054010 (2006).
%\bibitem{cnmel1} I. Vitev, Phys. Rev. C {\bf 75}, 064906 (2007).
%\bibitem{cnmel2} I. Vitev, A. Adil, H. van Hees, J. Phys. G {\bf 34}, S769-774 (2007).
\bibitem{thews} Thews, AIP Conf. Proc. 631 (2002) 490-524; arXiv:hep-ph/0206179v1 (2002)
\bibitem{ccbar} The ALICE Collaboration, J. High Energy Phys. {\bf 07}, 191 (2012).
%\bibitem{mats} T. Matsui and H. Satz, Phys. Lett. B {\bf 178}, 416 (1986). 
%\bibitem{Chu} M. C. Chu and T. Matsui, Phys. Rev. D {\bf 37}, 1851 (1988).
\bibitem{comov}  E. G. Ferreiro, Phys. Lett. B {\bf 731}, 57-63 (2014).
\bibitem{interpolate}  F. Bossu et al., arXiv:1103.2394v3 [nucl-ex], Apr (2012).
\bibitem{QQanni} Geoffrey T. Bodwin, Eric Braaten, G. Peter Lepage,  
%"Rigorous QCD analysis of inclusive annihilation and production of heavy quarkonium"
Phys. Rev. D {\bf 51}, 03 (1995).
\bibitem{gans} S. Ganesh, M. Mishra, Phys. Rev. C {\bf 88}, 044908 (2013).
\bibitem{woods} C. W. deJager, H. deVries, and C. deVries, Atomic Data and Nuclear Data Tables {\bf 14} 485, (1974).
%\bibitem{microbarn}  Vineet Kumar, Prashant Shukla, arXiv:1410.3299v1 [hep-ph], Oct (2014).
\bibitem{Madhu1} M. Mishra, C. P. Singh, V. J. Menon and Ritesh Kumar Dubey, Phys. Lett. B {\bf 656}, 45 (2007).
%\bibitem{Octet} C. Y. Wong, Phys. Rev. D {\bf 60}, 114025 (1999).
\bibitem{newvogt} V. Emelyanov, A. Khodinov, S. R. Klein, R. Vogt, Phys. Rev. Lett. {\bf81}, 1801-1804 (1998).
\bibitem{EPS09} K. J. Eskola, H. Paukkunen and C. A. Salgado, J. High Energy Phys. {\bf 0904}, 065 (2009).
%\bibitem{Laine1} M. Laine, O. Philipsen, M. Tassler and P. Romatschke, J. High Energy Phys. {\bf 03}, 054 (2007). 
%\bibitem{Laine2} M. Laine, O. Philipsen, M. Tassler, J. High Energy Phys. {\bf 09}, 066 (2007). 
%\bibitem{Laine3} M. Laine, Nucl. Phys. A {\bf 820}, 25c (2009).
%\bibitem{CMS2} The CMS Collaboration, Phys. Rev. Lett. {\bf 109}, 222301 (2012).
%\bibitem{PKSBAL} B. Alver et al., arXiv:0805.4411 [nucl-ex] (2008).
%\bibitem{PKSSSA} S. S. Adler et al., (PHENIX Collaboration), Phys. Rev. C {\bf 71}, 034908 (2005).
%\bibitem{PKSBJO} J. D. Bjorken, Phys. Rev. D {\bf 27}, 140 (1983).
%\bibitem{CMS} The CMS collaboration, J. High Energy Phys. {\bf 08}, 141 (2011); arXiv:1107.4800v2 [nucl-ex] (2011).
%\bibitem{abdul} A. Abdulsalam and Prashant Shukla, arXiv:1210.7584v1 [hep-ph] (2012).
%\bibitem{Temp} M. Strickland, Phys. Rev. Lett. {\bf 107}, 132301 (2011). 
%\bibitem{ALICE}  The ALICE Collaboration, Phys. Rev. Lett. {\bf 106}, 032301 (2011); arXiv:1012.1657v2 [nucl-ex] (2011).
%\bibitem{CMS3} The CMS Collaboration, Phys. Rev. Lett. {\bf 107}, 052302 (2011). 
\bibitem{CTEQ6} J. Pumplin, D. R. Stump, J. Huston, H. L. Lai, P. M. Nadolsky and W. K. Tung, JHEP {\bf 0207}, 012 (2002); arXiv:hep-ph/0201195.
\bibitem{tempform} Martin Shroedter, Robert L. Thews, and Johann Rafelski, Phys. Rev. C {\bf62}, 024905 (2000).
\bibitem{ptrelate} Jens Wiechula, 
%nuclear modification of $J/\psi$ production in Pb-Pb collisions at $\sqrt(S_{NN}) = 2.76TeV$
%Nuclear Physics A 00 (2012)
Nuclear Physics A {\bf 910-911}, 219, (2013)


%\bibitem{PKSPKS1} P. K. Srivastava and C. P. Singh. Phys. Rev. D {\bf 85}, 114016 (2012).
%\bibitem{PKS2} P. K. Srivastava and C. P. Singh, Phys. Rev. D {\bf 85}, 114016 (2012).
%\bibitem{PKS1} P. K. Srivastava, S. K. Tiwari and C. P. Singh, Phys. Rev. D {\bf 82}, 014023 (2010).

\end{thebibliography}
\end{document}